\documentclass[journal,10pt]{IEEEtran}
\usepackage[T1]{fontenc}

\usepackage{amsmath}
\usepackage{amssymb}
\usepackage{amsfonts}
\usepackage{graphicx}
\usepackage{epsfig}
\usepackage{psfrag}
\usepackage{cite}
\usepackage{latexsym}
\usepackage{lettrine}
\usepackage{url}
\usepackage{color}
\usepackage{multirow}
\usepackage{algorithmic}
\usepackage{algorithm}
\usepackage{subcaption}
\usepackage[font=footnotesize]{caption} 

\usepackage{verbatim}

\usepackage{stfloats}
\usepackage[english]{babel}
\usepackage{amsthm}
\usepackage{titlesec}

\usepackage{bm}
\usepackage{array}
\usepackage{hyperref}					
\hypersetup{colorlinks,
	linkcolor=blue,%
	anchorcolor=blue,
    citecolor=blue}

\newcommand{\Expct}[1]{\mathbb{E}\left\{#1\right\}}
\newcommand{\Var}[1]{\mathbb{D}\left\{#1\right\}}
\newcommand{\Abs}[1]{\left|#1\right|}
\newcommand{\Norm}[1]{\left\Vert#1\right\Vert}
\newcommand{\Real}[1]{\Re\left\{#1\right\}}
\newcommand{\Imag}[1]{\Im\left\{#1\right\}}
\newcommand{\Brkts}[1]{\left(#1\right)}

\newtheoremstyle{boldnote} 
  {3pt} 
  {3pt} 
  {\itshape} 
  {} 
  {\bfseries} 
  {.} 
  {.5em} 
  {\thmname{#1}\thmnumber{ #2}\thmnote{\bfseries{ (#3)}}} 

\theoremstyle{boldnote}
\newtheorem{assumption}{Assumption}

\theoremstyle{boldnote}\newtheorem{proposition}{\textbf{Proposition}}

\begin{document}
\title{Pulse Shaping for Random ISAC Signals: \\The Ambiguity Function Between Symbols Matters}
\author{
Zihan Liao, Fan Liu,~\IEEEmembership{Senior Member,~IEEE}, Shuangyang Li,~\IEEEmembership{Member,~IEEE}, \\Yifeng Xiong,~\IEEEmembership{Member,~IEEE}, Weijie Yuan,~\IEEEmembership{Senior Member,~IEEE},\\ Christos Masouros,~\IEEEmembership{Fellow,~IEEE}, and Marco Lops,~\IEEEmembership{Fellow,~IEEE}
\vspace{-10pt}

\thanks{Z. Liao is with the Department of Electrical and Electronic Engineering, F. Liu (corresponding author) and W. Yuan are with the School of System Design and Intelligent Manufacturing, Southern University of Science and Technology, Shenzhen, China. (email: liaozh2020@mail.sustech.edu.cn, liuf6@sustech.edu.cn, yuanwj@sustech.edu.cn).}
\thanks{S. Li is with the Faculty of Electrical Engineering and Computer Science, Technical University of Berlin, Berlin, 10587, Germany (e-mail: shuangyang.li@tu-berlin.de).}
\thanks{Y. Xiong is with the School of Information and Electronic Engineering,
Beijing University of Posts and Telecommunications, Beijing 100876, China. (e-mail: yifengxiong@bupt.edu.cn)}
\thanks{C. Masouros is with the Department of Electronic and Electrical Engineering, University College London, London, WC1E 7JE, UK (e-mail: chris.masouros@ieee.org).}
\thanks{M. Lops is with the Department of Electrical and Information Technology, University of Naples Federico II, 80138 Naples, Italy, and also with Consorzio Nazionale Interuniversitario per le Telecomunicazioni, 43124 Parma, Italy (e-mail: lops@unina.it).}
\vspace{-10pt}
}

\maketitle
\begin{abstract}
Integrated sensing and communications (ISAC) has emerged as a pivotal enabling technology for next-generation wireless networks. Despite the distinct signal design requirements of sensing and communication (S\&C) systems, shifting the symbol-wise pulse shaping (SWiPS) framework from communication-only systems to ISAC poses significant challenges in signal design and processing This paper addresses these challenges by examining the ambiguity function (AF) of the SWiPS ISAC signal and introducing a novel pulse shaping design for single-carrier ISAC transmission. We formulate optimization problems to minimize the average integrated sidelobe level (ISL) of the AF, as well as the weighted ISL (WISL) while satisfying inter-symbol interference (ISI), out-of-band emission (OOBE), and power constraints. Our contributions include establishing the relationship between the AFs of both the random data symbols and signaling pulses, analyzing the statistical characteristics of the AF, and developing algorithmic frameworks for pulse shaping optimization using successive convex approximation (SCA) and alternating direction method of multipliers (ADMM) approaches. Numerical results are provided to validate our theoretical analysis, which demonstrate significant performance improvements in the proposed SWiPS design compared to the root-raised cosine (RRC) pulse shaping for conventional communication systems.
\end{abstract}

\begin{IEEEkeywords}
Deterministic-random tradeoff, pulse shaping, integrated sensing and communications (ISAC).
\end{IEEEkeywords}
\vspace{-10pt}

\section{Introduction}

\IEEEPARstart{I}{ntegrated} sensing and communications (ISAC) has emerged as a pivotal enabling technology for next-generation wireless networks, such as 5G-advanced and 6G \cite{chafii2023twelve}. This technology seeks for profound integration between wireless sensing and communication (S\&C) to facilitate the co-design of both functionalities, thereby enhancing hardware, spectral, and energy efficiency while obtaining mutual performance gains \cite{pin2021integrated}\cite{liu2020joint}. As a result, ISAC is well-recognized as a promising technology for 6G wireless networks, which is capable of supporting a variety of emerging applications including intelligent transportation, smart manufacturing, and environment monitoring \cite{liu2022integrated}.

To deploy ISAC in the next-generation wireless networks, a feasible low-cost solution is to directly utilize the standardized communication waveform for sensing \cite{wei2023integrated,cui20225g,li2024frame}. Such communication-centric schemes typically exploit physical layer reference signals (i.e., pilots) as sensing signals, including demodulation reference signal (DMRS), phase tracking reference signal (PTRS), channel state information reference signal (CSI-RS), and positioning reference signal (PRS), among others \cite{wei20225g}. These reference signals, however, occupy only 10\% of the time-frequency resources in a typical communication frame structure, resulting in low range and Doppler resolution \cite{lin20215g}. Towards that end, one has to reuse the remaining 90\% of the resources occupied by communication data payload for target sensing, thus improving the sensing performance \cite{saad2019vision}. 

Nevertheless, S\&C systems exhibit distinct signal design pipelines and thus differ in their nature of optimality \cite{xiong2023fundamental}, which impose challenges in reusing data payload signals for sensing. To be more specific, communication systems typically employ SWiPS at the transmitter, and perform symbol-wise matched filtering at the receiver, in the hope to eliminate the ISI and constrain the signal bandwidth through precisely controlling the shape of each individual symbol \cite{gandhi2013implementation,xia1997family}. In contrast to that, radar systems do not need to remove the ISI as there is no information symbol to be conveyed. Instead, they focus on optimizing the overall ambiguity characteristics of the signal without optimizing the shape of each fast-time sub-pulse \cite{sturm2011waveform}\cite{richards2010principles}. More importantly, communication data modulated over the traditional communication-only SWiPS signals are \textit{random}, where higher degree of randomness indicates greater information-carrying capabilities \cite{neeser1993proper}. Provably, such randomness may degrade the radar sensing performance, since conventional sensing signals are usually meticulously designed deterministic signals with favorable ambiguity properties, (e.g. linear frequency modulation signals) \cite{levanon2004radar}.


The above distinctions between S\&C may be further interpreted by a pair of tradeoffs: the deterministic-random tradeoff (DRT) and the subspace tradeoff (ST) \cite{xiong2024torch}. From the perspective of DRT, the inherent randomness of communication data introduces variability in the signal, leading to fluctuations in sensing performance metrics such as the ambiguity function (AF) \cite{xiao2022waveform,lu2023random}. This variability may reduce the ranging accuracy and probability of detecting multiple targets, introduce ghost targets, and increase the false alarm rate \cite{xiao2022waveform}. From the perspective of ST, conventional SWiPS designs focus on communication-favorable signal subspaces, neglecting sensing-related properties. Specifically, traditional SWiPS requires certain discrete points of the AF to be zero (corresponding to the Nyquist criterion) \cite{Baas2004Pulse}, as will be demonstrated in later sections. On the contrary, sensing tasks usually require maintaining a low sidelobe level across a continuous delay-Doppler region of interest, thus to guarantee effective target estimation and detection performance \cite{vizitiu2014some}.

In light of the above two types of tradeoff, achieving the optimal S\&C performance neccesitates a thorough re-evaluation of the constraints and objectives involved to tailor to ISAC SWiPS signaling. A significant number of works on ISAC signal design were motivated from discrete models formulated by precoded/unprecoded communication symbols \cite{liu2021cramer,xiao2022waveform,wei2023waveform}. This is valid for a communication system since its end-to-end input-output relation can be naturally modeled as discrete. However, working on discrete signal models may not fully characterize the sensing performance, due to the fact that one may lose critical target information located in between communication symbols. Therefore, it is essential to thoroughly analyze and optimize the sensing performance by taking the pulse shaping design into account. In the SWiPS regime, the symbol sequence is convolved with a pulse-shaping filter, such that each symbol is associated with a continuous signaling pulse adapting to various requirements of the system \cite{Baas2004Pulse}. Typical constraints imposed on the SWiPS filters include the OOBE \cite{matthe2014influence} and ISI constraints, where the zero-ISI constraint corresponds to the Nyquist criterion \cite{Vahlin1996Optimal}. Despite their effect of ensuring the communication performance, these constraints may not be necessarily beneficial for the sensing functionality, especially for implementing sensing using random data signals. This motivates us to conceive an SWiPS filter by incorporating ambiguity properties of the random ISAC signals in addition to communication-oriented constraints. To that aim, two main challenges must be addressed: 1) Modeling the AF within the SWiPS framework to incorporate sensing-oriented objectives into SWiPS signaling, accounting for the randomness introduced by communication symbols. 2) Formulating and solving the SWiPS design problem, which, upon modeling the ambiguity characteristics of the SWiPS random signal, aims to reduce the sidelobe level of the AF while ensuring that the communication constraints are met.

The aforementioned challenges complicate the analysis and optimization under conventional sensing frameworks. Against this background, in this paper, we examine the statistical properties of the AF of the SWiPS signal and introduce a novel SWiPS design for single-carrier ISAC that may enhance the range-Doppler detection performance. We formulate optimization problems for the ISAC SWiPS filter design, in an effort to reduce the average integrated sidelobe level (ISL) as well as the weighted ISL (WISL) while adhering to the communication ISI, OOBE, and power constraints. For clarity, our contributions are summarized as follows:
\begin{itemize}
    \item We establish the relationship between the AF of the ISAC signal and that of the signaling pulse. Subsequently, we investigate the statistical attributes of the AF for SWiPS signals modulated with random i.i.d. communication symbols, where the expectation and variance of the AF may be explicitly expressed in closed form.
    \item We define the ISL for evaluating the quality of the AF. To accommodate the SWiPS design for broader ISAC scenarios where certain prior information of the environment is available, we present an advanced sensing input-output model concerned with channel scattering information, and define the WISL as a more generic performance metric.
    \item We formulate the SWiPS design optimization problem to minimize the WISL while satisfying the ISI, OOBE, and power constraints, which is non-convex in general. As a special case, we demonstrate that by restricting the region of interest of the AF to the zero-Doppler slice (thereby reducing the AF to the auto-correlation function (ACF)), the SWiPS design problem may be reformulated as a convex quadratic programming (QP).
    \item We develop an algorithmic framework utilizing the SCA method to address the non-convex WISL minimization problem, which transforms the problem into solvable convex sub-problems with linearly approximated objective functions. Additionally, to efficiently solve the convex ACF WISL minimization problem, we devise an algorithmic framework employing the alternating direction method of multipliers (ADMM).
\end{itemize}
Numerical results validate the accuracy of the statistical analysis of the AF. Additionally, these results demonstrate that the proposed SWiPS design significantly reduces the ranging sidelobe levels, as well as both the ISL and WISL, compared to traditional RRC pulse shaping, while maintaining the communication bit rate and desired signal properties.

The remainder of this paper is structured as follows. Section II introduces the system model and performance metrics for S\&C. Section III analyzes the statistics of the AF. Section IV presents a case study, introducing the ISL and WISL minimization problems and analyzing the S\&C performance tradeoff for our benchmark, the RRC. Section V presents the algorithms for solving the WISL minimization problems. Simulation results are presented in Section VI, and conclusions are provided in Section VII. Lastly, proofs of the propositions in the paper are given in Appendices A to C.

\textbf{Notation:} In this paper, $\mathbf{A}$, $\mathbf{a}$, and $a$ denote a matrix, vector, and scalar, respectively. We use $\Real{\cdot}$, $\Imag{\cdot}$, $\Expct{\cdot}$, $|\cdot|$, $\|\cdot\|$, $(\cdot)^\top$, $(\cdot)^*$, $(\cdot)^H$, $\left\lceil\cdot\right\rceil$, $\left[\cdot\right]$, $\mathbf{1}_N$, $\mathbf{I}_{N}$, and $\mathbf{0}_{N,M}$ to denote the real part and imaginary of a complex number, expectation, modulus of a complex number, Frobenius norm, transpose, conjugate, Hermitian, round, ceil, identity vector of size $N \times 1$, identity matrix of size $N \times N$, and zero matrix of size $N \times M$ respectively.

\section{Signal Model and Performance Metrics}

\subsection{SWiPS ISAC Signal Model}

Within the structure of SWiPS, the symbol at the $n$-th position in a communication frame of length $L$, denoted as $s_n$, undergoes modulation to create a symbol pulse $s_n(t) = s_n g(t)$.
Here, $g(t)$ represents the impulse response of the pulse shaping filter. Subsequently, the ISAC signal can be articulated as the aggregate of the time-shifted symbol pulses as below
\begin{equation}
    s(t) = \sum_{n=0}^{L-1} s_n(t-nT) = \sum_{n=0}^{L-1} s_n g(t - nT),
\end{equation}
where $T$ is the symbol duration.

Before elaborating on the metrics of S\&C, we establish the following assumptions which are commonly made in practical communication systems:
\begin{assumption}[Circularly Symmetric Constellation]
\label{assump:symmetric}
In this paper, we focus on constellations that are considered proper, or circularly symmetric, characterized by the condition that the pseudo-correlation \( \Expct{s_n s_m} = 0 \) for all \( n, m \), and that the constellation center is at the origin, resulting in \( \Expct{s_n} = 0 \) for all \( n \). \\
\underline{Remark:} This assumption holds for constellations where the distribution in one quadrant can be rotated to match the distribution in any other quadrant. This assumption is satisfied by common types of constellations, including all the PSK and QAM constellations except for BPSK and 8-QAM.
\end{assumption}

\begin{assumption}[Independent Symbols]
\label{assump:independent}
We assume that $s_n$ and $s_m$ are independent when $n$ does not equal to $m$, leading to $\Expct{s_n s_m^*}=0, \forall n\neq m$.
\end{assumption}

\begin{assumption}[Identical Constellation]
\label{assump:identical}
We assume that all symbols within a single frame share the same constellation, whose energy is normalized to $1$, implying that $\Expct{\Abs{s_n}^2}=1, \; \Expct{\Abs{s_n}^4}=\mu_4, \forall n$,
where $\mu_4\geq 1$ is the kurtosis of the constellation.
\end{assumption}

\subsection{Sensing Metric: Ambiguity Function}

Let us first define the ambiguity function (AF) of $s(t)$ by
\begin{equation}
\begin{aligned}
\chi(\tau, \nu) &= \int s(t) s^*(t-\tau)e^{-j2\pi\nu t} dt,
\end{aligned}
\end{equation}
which denotes the correlation between the signal $s(t)$ and its time- and frequency-shifted counterpart $s(t-\tau)e^{j2\pi\nu t}$. Given that the frame signal consists of symbol pulses, the AF of the frame signal can be expressed as the sum of the cross AF between the time-shifted symbol pulses. Through basic algebraic manipulation, it can be expressed as follows:
\begin{equation}
\chi(\tau, \nu) = \sum_{n=0}^{L-1} \sum_{m=0}^{L-1} s_n s_m^* \psi_{n,m}(\tau,\nu),
\end{equation}
where $\psi_{n,m}(\tau,\nu)=e^{-j 2 \pi n \nu T} \psi(\tau+(m-n) T, \nu)$ represents the cross-ambiguity function between the $n$-th and $m$-th time-shifted pulses, and $\psi(\tau, \nu) = \int g(t) g^*(t-\tau)e^{-j2\pi\nu t} dt$ is the AF of $g(t)$.

The zero-Doppler slice of the AF, which is equivalent to the ACF of $s(t)$, may be expressed as
\begin{equation}
\label{eq:acf_s_from_g}
\chi(\tau, 0) =\sum_{n=0}^{L-1} \sum_{m=0}^{L-1} s_n s_m^* G(\tau+(m-n) T),
\end{equation}
where $G(\tau) = \psi(\tau,0)$ is the ACF of $g(t)$. Let $U(f)$ represent the Fourier transform of $g(t)$. The AF of $g(t)$ may be alternatively expressed in the frequency domain as
\begin{equation}
    \psi(\tau, \nu) = \int U(f) U^*(f-\nu)e^{j2\pi f\tau} df.
\end{equation}
Accordingly, the Zero-Doppler Slice, namely, the ACF, may be given by
\begin{equation}
\label{eq:acf_spectr}
    G(\tau) = \psi(\tau, 0) = \int \omega(f) e^{j2\pi f\tau} df,
\end{equation}
where $\omega(f)=|U(f)|^2$ is the Energy Spectral Density (ESD) of $g(t)$.
Eq. \eqref{eq:acf_spectr} simply suggests that the ACF is equivalent to the inverse Fourier transform of the ESD, following the Wiener–Khinchin theorem. This indicates that $G(\tau)$ can be represented as a linear function of $\omega(f)$. Moreover, as seen from \eqref{eq:acf_s_from_g}, since $\chi(\tau, 0)$ is a linear combination of $G(\tau)$ for different $\tau$, it becomes clear that $\chi(\tau, 0)$ can also be expressed linearly in terms of $\omega(f)$. Consequently, this allows us to formulate the pulse shaping design problem as a convex programming when taking into account the ACF only, as will be detailed in later sections.

\subsection{Communicaion Metrics: OOBE and ISI}
\label{sec:OOBE_isi}

\subsubsection{OOBE Constraint}
The OOBE constraint aims to control the energy leakage of $s(t)$, which could be realized by regulating the OOBE leakage of $g(t)$. Therefore, we define the OOBE constraint such that the energy of $g(t)$ outside the desired bandwidth must be less than $\varepsilon_{OB}$, expressed as
\begin{equation}
    \mathcal{C}_{\varepsilon}^{OOBE}: \int_{-\infty}^{-B/2} \omega(f) df + \int_{B/2}^{\infty} \omega(f) df \leq \varepsilon_{OB},
\end{equation}
where $B$ refers to the bandwidth of the pulse.
$\varepsilon_{OB}$ must be kept minimal to prevent interference between adjacent bands. In the extreme case, one may assign $\varepsilon_{OB}=0$, resulting in the zero-leakage constraint $\mathcal{C}_{0}^{OOBE}$.

\subsubsection{ISI Constraint}

Considering an additive white Gaussian noise (AWGN) channel, the representation of the matched filtered signal at the receiver side is given by
\begin{equation}
\begin{aligned}
    y(t) &= s(t)*g(-t) + \tilde{n}(t) = \sum_{n=0}^{L-1} s_n G(t - nT) + \tilde{n}(t),
\end{aligned}
\end{equation}
where $\tilde{n}(t)$ refers to the noise part of the matched filter output. After sampling at $t=kT,k=0,1,\cdots,L-1$, we have
\begin{equation}
    y_k = h_c\sum_{n=0}^{L-1} s_n G((k - n)T) + \tilde{n}(kT),
\end{equation}
which represents the communication input-output relationship.
Hence, to keep the ISI within an acceptable range, the following constraint must be met:
\begin{equation}
    \mathcal{C}_{\varepsilon}^{ISI}: |G(nT)|^2\leq\varepsilon_{ISI}, n=\cdots,-2,-1,1,2,\cdots.
\end{equation}

Alternatively, to meet the Nyquist criterion and eliminate ISI, the $\varepsilon_{ISI}$ should be zero.
According to \cite{barry2012digital}, this time-domain Nyquist constraint is equivalent to 
\begin{equation}
    \sum_{m=-\infty}^{\infty} \omega(f+m/T) = c_T,
\end{equation}
where $c_T$ is a constant. This is also known as the constant folded spectrum criterion. Let us denote the roll-off factor of the pulses as $\beta\in[0,1]$, which links the symbol duration $T$ to the bandwidth $B$ via $BT=1+\beta$. Consequently, we have $\frac{1}{2T}\leq\frac{B}{2}\leq\frac{1}{T}$. With $\mathcal{C}_{0}^{OOBE}$ satisfied, the Nyquist condition can be recast as
\begin{equation}
\label{eq:freq_no_isi_constr}
\mathcal{C}_{0}^{ISI}:
\left\{
\begin{aligned}
&\omega(f) = c_T, &&0\leq f\leq\frac{1}{T}-\frac{B}{2}, \\
&\omega(f) + \omega\left(\frac{1}{T}-f\right) = c_T, &&\frac{1}{T}-\frac{B}{2}\leq f\leq\frac{B}{2}.
\end{aligned}
\right.
\end{equation}

Additionally, the pulses must adhere to the energy constraint. We assume the energy budget is $E_g$, leading to the following constraint:
\begin{equation}
    \mathcal{C}^E: \int \Abs{g(t)}^2 dt \leq E_g \text{ or } \int \omega(f) df \leq E_g.
\end{equation}
Notice that the energy constraint imposed on $\omega(f)$ may also be implicitly guaranteed in $\mathcal{C}_{0}^{ISI}$, where there would be a linear relationship between $c_T$ and $E_g$.

For simplicity, in the following section, we use the notation $S_g$ to refer to the set of $g(t)$ that meets the constraints $\mathcal{C}_{\varepsilon}^{OOBE}$, $\mathcal{C}_{\varepsilon}^{ISI}$, and $\mathcal{C}^E$. Similarly, we define $\mathcal{S}_{\omega}$ as the set of $\omega(f)$ that fulfills the constraints $\mathcal{C}_{0}^{OOBE}$, $\mathcal{C}_{0}^{ISI}$ and $\mathcal{C}^E$. We highlight here that the set $\mathcal{S}_g$ is non-convex in terms of $g(t)$, whereas $\mathcal{S}_{\omega}$ is a linear convex set with respect to $\omega(f)$.

\section{Statistics of the Ambiguity Function}

\subsection{Two Parts of the AF}

To derive the statistical properties of the AF, we first split $\chi(\tau, \nu)$ to two parts by
\begin{equation}
\chi(\tau, \nu) = \underbrace{\sum_{n=0}^{L-1}\left|s_n\right|^2 \psi_{n,n}(\tau,\nu)}_{\chi_s(\tau, \nu)}  +\underbrace{\sum_{n=0}^{L-1} \sum_{\underset{m \neq n}{m=0,}}^{L-1} s_n s_m^* \psi_{n,m}(\tau,\nu)}_{\chi_c(\tau, \nu)}.
\label{eq:twopart}
\end{equation}
$\chi_s(\tau, \nu)$ describes the part contributed by the self-ambiguity of the symbols, while $\chi_c(\tau, \nu)$ describes the part contributed by the cross-ambiguity between different symbols. By doing so, one may arrive at the following propositions:

\begin{proposition}
\label{prop:chi}
The expectations of $\chi_s(\tau, \nu)$ and $\chi_c(\tau, \nu)$ have the following properties:
\begin{align}
    &\Expct{\chi_s(\tau,\nu)}=\sum_{n=0}^{L-1} e^{j 2 \pi n \nu T} \psi(\tau, \nu), \label{eq:chis_expct}\\
    &\Expct{\chi_c(\tau,\nu)}=0, \label{eq:zero_chic}\\
    &\Expct{\chi_s(\tau,\nu)\chi_c^*(\tau,\nu)}=0, \label{eq:zero_corr_sc}
\end{align}
\begin{proof}
    See Appendix \ref{sec:app_B}.
\end{proof}
\end{proposition}

\begin{proposition}
\label{prop:chisc_var}
The variances of $\chi_s(\tau, \nu)$ and $\chi_c(\tau, \nu)$ can be expressed as
\begin{align}
    &\Var{\chi_s(\tau, \nu)} = L(\mu_4-1)|\psi(\tau,\nu)|^2, \label{eq:chis_var} \\
    &\Var{\chi_c(\tau, \nu)} = \sum_{|n|<L,n\neq 0} (L-|n|) |\psi(\tau+nT, \nu)|^2, \label{eq:chic_var} \\
    &\Var{\chi(\tau,\nu)} = \Var{\chi_s(\tau, \nu)} + \Var{\chi_c(\tau, \nu)}. \label{eq:var_addition}
\end{align}
\begin{proof}
    See Appendix \ref{sec:app_C}.
\end{proof}
\end{proposition}
We may now present the expectation and variance of the AF in the below subsections.

\subsection{Expectation of the Ambiguity Function}

According to \eqref{eq:twopart} and \eqref{eq:zero_chic}, the expectation of the $\chi(\tau, \nu)$ can be written as 
\begin{equation}
\begin{aligned}
\mathbb{E}\{\chi(\tau, \nu)\} &= \mathbb{E}\{\chi_s(\tau, \nu)\} = \psi(\tau, \nu)\sum_{n=0}^{L-1} e^{j 2 \pi n \nu T}.
\end{aligned}
\end{equation}

\subsection{Variance of the Ambiguity Function}

According to Proposition \ref{prop:chisc_var}, we may express the variance of the AF $\chi(\tau, \nu)$ as
\begin{equation}
\begin{aligned}
&\Var{\chi(\tau,\nu)} = \sum_{|n|<L} \tilde{\alpha}_n |\psi(\tau+nT, \nu)|^2,
\end{aligned}
\end{equation}
where $\tilde{\alpha}_n = L(\mu_4 - 1)$ when $n = 0$, and $L - |n|$ otherwise. With the expectation and the variance of $\chi(\tau, \nu)$ at hand, it is viable to compute the expectation of the squared AF (SAF) in the form of
\begin{equation}
\label{eq:theo_saf}
\begin{aligned}
\mathbb{E}\left\{|\chi(\tau, \nu)|^2\right\} &= \mathbb{D}\{\chi(\tau, v)\}+|\mathbb{E}\{\chi(\tau, v)\}|^2 \\
&=  \sum_{|n|<L} \alpha_n(\nu) \Abs{\psi(\tau+nT, \nu)}^2
\end{aligned}
\end{equation}
where 
\begin{equation}
    \alpha_n(\nu)=
    \left\{
    \begin{aligned}
        &L(\mu_4-1)+\left|\sum_{m=0}^{L-1}e^{j2\pi m\nu T}\right|^2, &&n=0, \\
        &L-|n|, &&n\neq 0,|n|<L.
    \end{aligned}\right.
\end{equation}
Without the loss of generality, in this paper, we constrain the energy of the pulse $E_g$ to be $1$, and assume that the ISI is sufficiently small, which implies that $\Abs{\psi(\tau+nT, \nu)}^2$ is small enough for non-zero $n$. Consequently, the squared energy of $s(t)$, namely, $\mathbb{E}\left\{|\chi(0, 0)|^2\right\}$, may be approximated by
\begin{equation}
    \mathbb{E}\left\{|\chi(0,0)|^2\right\} \approx \alpha_0(0) =  L(\mu_4-1) + L^2.
\end{equation}
This term is employed to normalize the AF in the later sections and the validity of the approximation is demonstrated by our numerical results in Figure \ref{fig:safs}. 

\section{Case Study}

\subsection{ISL Minimization}
Given the randomness of the transmitted symbol sequence, it would be more appropriate to consider the expectation of ISL of $\chi(\tau,\nu)$ rather than the ISL of a single instance of the symbol sequence. By leveraging the results in preceding sections, the expectation of the ISL may be given as
\begin{equation}
\begin{aligned}
\mathcal{L}_{ISL}\{g(t)\}&=\mathbb{E}\left\{\iint_{\odot}|\chi(\tau, \nu)|^2 d \tau d \nu\right\} \\
&= \iint_{\odot}\mathbb{E}\left\{|\chi(\tau, \nu)|^2\right\} d\tau d\nu \\
&= \sum_{|n|<L} \iint_{\odot}\alpha_n(\nu)|\psi(\tau+nT, \nu)|^2 d\tau d\nu,
\end{aligned}
\label{eq:islr}
\end{equation}
where $\odot$ represents the delay-Doppler region of interest. The pulse shaping filter should thus be designed by solving the following problem
\begin{equation}\label{problem_1}
    \min_{g(t)} \; \mathcal{L}_{ISL}\{g(t)\} \; s.t.\; g(t)\in\mathcal{S}_g.
\end{equation}

\subsection{WISL Minimization in the Presence of Ground Clutter}

In typical situations, the received sensing signal is reflected not only from the targets but also from interfering environmental objects, e.g., clutter reflected from the ground, which is range-spread in general. To reduce the clutter interference, one may exploit certain prior information about the statistics of the scattering environment in the pulse shaping design. To that end, a more comprehensive sensing signal model that incorporates environmental information is presented below.

A range- and Doppler-spread channel responds to the signal \( s(t) \) by returning the continuous superposition:
\begin{equation}
    x(t) = \iint \alpha(\lambda, \mu)s(t - \lambda)e^{j2\pi\mu t} \, d\lambda \, d\mu,
\end{equation}
where \( \alpha(\lambda, \mu) \) is a zero-mean random field—typically modeled as complex Gaussian—encapsulating the scattering properties of the environment and the targets.

Suppose \( \sigma(\lambda_1, \lambda_2, \mu_1, \mu_2) = \mathbb{E}[\alpha(\lambda_1, \mu_1)\alpha^*(\lambda_2, \mu_2)] \) represents the channel scattering information (CSI) of the environment, following the Wide-Sense Stationary Uncorrelated Scattering (WSSUS) model \cite{sadowsky1998correlation}, which implies:
\begin{equation}
    \sigma(\lambda_1, \lambda_2, \mu_1, \mu_2) = \sigma(\lambda_1, \mu_1)\delta(\lambda_1 - \lambda_2)\delta(\mu_1 - \mu_2).
\end{equation}
Let us consider the output signal when \( x(t) \) is matched filtered through \( s^*(t - \tau)e^{-j2\pi\nu t} \), yielding
\begin{equation}
\begin{aligned}
    X(\tau, \nu) &= \int s^*(t - \tau)e^{-j2\pi\nu t}x(t) \, dt \\
    &= \iint \alpha(\lambda, \mu)\chi(\tau - \lambda, \nu - \mu) \, d\lambda \, d\mu.
\end{aligned}
\end{equation}

\begin{proposition}
\label{prop:wssus}
The mean square value of \( X(0, 0) \), under the WSSUS assumption, is
\begin{equation}
    \mathbb{E}[|X(0,0)|^2] = \iint \sigma(\lambda, \mu)\mathbb{E}[|\chi(\lambda, \mu)|^2] \, d\lambda \, d\mu.
\end{equation}
\begin{proof}
See Appendix \ref{sec:app_E}.
\end{proof}
\end{proposition}

Suppose that we are detecting point-like targets in the presence of the clutter. The channel scattering function, under these circumstances, takes the form $\sigma(\lambda, \mu) = \sigma_T\delta(\lambda)\delta(\mu) + \sigma_c(\lambda, \mu)$, where \( \sigma_c(\lambda, \mu) \) represents the scattering function of diffuse clutter. Consequently, we have:
\begin{equation}
\begin{aligned}
    \mathbb{E}[|X(0,0)|^2] &= \sigma_T\mathbb{E}[|\chi(0,0)|^2] \\
    &\quad + \iint \sigma_c(\lambda, \mu)\mathbb{E}[|\chi(\lambda, \mu)|^2] \, d\lambda \, d\mu.
\end{aligned}
\end{equation}

In this representation, the component of the signal reflected from the target corresponds to the first term, while the segment pertaining to undesired reflection interference is associated with the second term. The second term can be interpreted as a WISL, where \( \sigma_c(\lambda, \mu) \) quantifies the interference from an object deviating by \( (\lambda, \mu) \) in the range-Doppler space. For brevity, we use the notation:
\begin{equation}
\begin{aligned}
    &\mathcal{L}_{WISL}\{g(t)\} = \iint \sigma_c(\tau, \nu)\mathbb{E}[|\chi(\tau, \nu)|^2] \, d\tau \, d\nu \\
    &\quad= \sum_{|n|<L} \iint_{\odot} \alpha_n(\nu)\sigma_c(\tau, \nu)|\psi(\tau+nT, \nu)|^2 \, d\tau \, d\nu.
\end{aligned}
\end{equation}
It is evident that \( \mathcal{L}_{ISL}\{g(t)\} \) is a special case of \( \mathcal{L}_{WISL}\{g(t)\} \), where \( \sigma_c(\tau, \nu) = 1 \) for all \( (\tau, \nu) \in \odot \) and \( \sigma_c(\tau, \nu) = 0 \) elsewhere.

In such a case, the pulse shaping filter should be conceived through solving the following problem:
\begin{equation}\label{problem_2}
    \min_{g(t)} \; \mathcal{L}_{WISL}\{g(t)\} \; s.t. \; g(t) \in \mathcal{S}_g.
\end{equation}
In the case that the environmental scattering information is not available, a prudent choice is to set \( \sigma_c(\tau, \nu) = 1 \) in the region of interest, which reduces to the ISL minimization problem.

\subsection{ISL/WISL Minimization: The ACF Case}

As discussed in Section \ref{sec:OOBE_isi}, if we define \( g(t) \) as a band-limited pulse—thereby ensuring zero OOBE energy leakage—the Nyquist criterion for zero ISI may be expressed as a linear constraint in terms of \(\omega(f)\). By focusing only on the zero-Doppler slice, the WISL, which is the sum of squares of \(\chi(\tau,0)\) in this context, can be represented as a linear function of \(\omega(f)\). Recall that in \eqref{eq:acf_spectr} we show that $\psi(\tau,0)$ is the linear transform of $|U(f)|^2$, rendering $|\psi(\tau,0)|^2$ as a convex quadratic form of $\omega(f)$. Then $\mathcal{L}_{ISL}\{g(t)\}$ and $\mathcal{L}_{WISL}\{g(t)\}$, which are the weighted integral of $|\psi(\tau,0)|^2$, are convex quadratic forms of $\omega(f)$ as well. Accordingly, the SWiPS WISL optimization problem under Nyquist criterion and zero OOBE constraint can be formulated as a convex QP problem with linear constraints. This formulation allows for the development of efficient algorithms to solve the problem. Furthermore, as will be demonstrated in Section \ref{sec:results}, the improvement in WISL for the Doppler slice under the SWiPS scenario is marginal. Therefore, optimizing WISL for the ACF may be a more practical option.

Following the above discussion, we can rewrite the WISL optimization for the ACF subject to the Nyquist criterion and zero OOBE constraint as 
\begin{align}
    &\min_{\omega(f)} \; \mathcal{L}_{WISL}\{\omega(f)\} \; s.t.\; \omega(f)\in\mathcal{S}_{\omega}, \label{opt:wisl_acf}
\end{align}
Apparently, \eqref{opt:wisl_acf} is a convex QP problem with linear constraints. In the later sections, we refer to this problem as the Nyquist-ACF-QP problem.

\subsection{S\&C Performance Tradeoff of SWiPS}

In this subsection, we examine the S\&C tradeoff by analyzing the ACF of the ISAC signal, under the condition that $L$ is adequately substantial, and the ISI is small enough to be disregarded.

\subsubsection{Asymptotic Behavior of $\chi(\tau,0)$}
Let us first investigate the asymptotic behavior of $\chi(\tau,0)$ when $L \to \infty$.
\begin{equation}
\begin{aligned}
&\lim_{L\rightarrow\infty} \frac{\mathbb{E}\left\{|\chi(\tau,0)|^2\right\}}{\mathbb{E}\left\{|\chi(0, 0)|^2\right\}}\approx\lim_{L\rightarrow\infty} \frac{\mathbb{E}\left\{|\chi(\tau,0)|^2\right\}}{\alpha_0(0)} \\
&= |\psi(\tau,0)|^2 \\
&\quad + \lim_{L\rightarrow\infty} \sum_{|n|<L,n\neq 0} \frac{L-|n|}{L(\mu_4-1) + L^2}|\psi(\tau+nT, 0)|^2 \\
& = |\psi(\tau,0)|^2
\end{aligned}
\end{equation}
This implies that when $L$ is sufficiently large, the normalized mean square of the ACF of the SWiPS ISAC signal would approach $\psi(\tau,0)$, namely, the ACF of the pulse itself. Therefore, instead of considering $\chi(\tau,0)$ directly, one may also analyze $\psi(\tau,0)$ in the case of a large $L$. A particular family of pulses that is extensively utilized and meets the Nyquist criterion is the RRC pulse, whose ESD is given as
\begin{equation}
\begin{aligned}
    &\Abs{U_{rrc}(f)}^2 = \\
    &\begin{cases} 
        T, & 0 \leq |f| \leq \frac{1-\beta}{2T}, \\
        \frac{T}{2} \left[ 1 + \cos\left( \frac{\pi T}{\beta} \left( |f| - \frac{1-\beta}{2T} \right) \right) \right], & \frac{1-\beta}{2T} < |f| \leq \frac{1+\beta}{2T}, \\
        0, & |f| > \frac{1+\beta}{2T}.
    \end{cases}
\end{aligned}
\end{equation}

\subsubsection{Sidelobe Level and Symbol Rate of RRC}

\begin{figure}[t]
    \centering
    \includegraphics[width=0.85\linewidth]{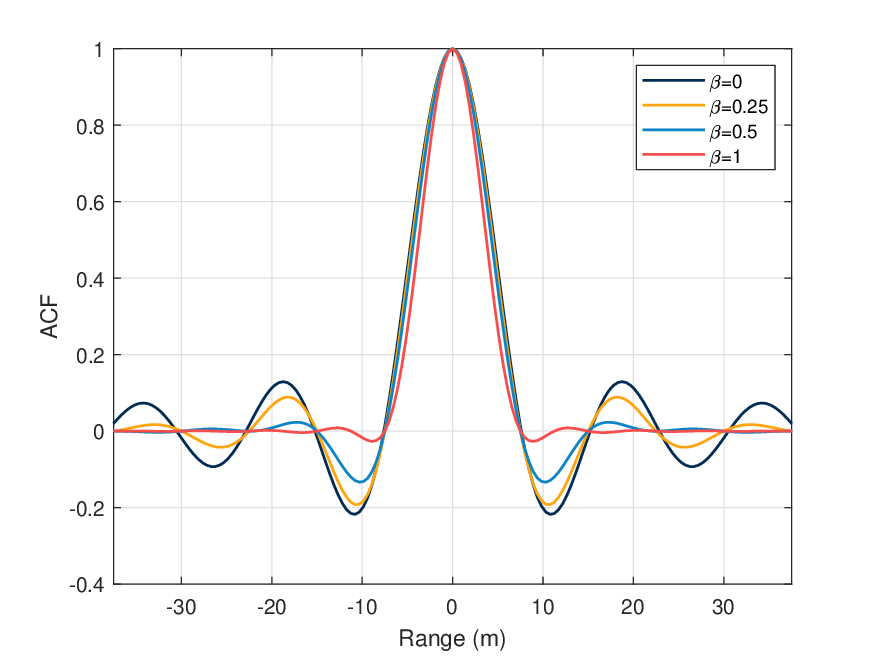}\vspace{-5pt}
    \caption{Plot of $G(\tau)$ with different $\beta$.}\vspace{-10pt}
    \label{fig:rrc_acf}
\end{figure}

The plot of the ACF of the RRC pulses with different $\beta$ is shown in Figure \ref{fig:rrc_acf}, which may be viewed as an approximation of the ACF of the random ISAC signal shaped by the RRC pulse for large $L$.
Obviously, as $\beta$ increases, the sidelobe level decreases.
The bandwidth of the RRC is given by $B=(\beta+1)/T$. Thus the symbol rate per Hz can be expressed as
\begin{equation}
    R_{\beta} = \frac{1}{TB} = \frac{1}{\beta+1}.
\end{equation}
Clearly, with an increase in $\beta$, there is a corresponding decrease in $R_{\beta}$, and concurrently, as previously stated, the sidelobe level also decreases. This demonstrates the tradeoff between S\&C performance, suggesting that adjusting $\beta$ may control the S\&C tradeoff.

\section{Pulse Shaping Optimization and Algorithms}
It is noteworthy that problems \eqref{problem_1}, \eqref{problem_2}, and \eqref{opt:wisl_acf} are continuous functional optimization problems, which could be highly computationally expensive to solve. In this section, we reformulate the optimization problems into discrete versions, and develop tailored algorithms to solve them.
\subsection{WISL Minimization: General Formulation}
\subsubsection{Discretization of the Pulse}
Let $g_k=g(kT_s)$ be a sampled sequence of $g(t)$ of length $L_g$ with sampling frequency established as $f_s$, which can be stacked into a vector $\mathbf{g}\in\mathbb{R}^{L_g}$. Accordingly, its frequency spectrum may be discretized as $U_k=U(k/L_gT_s)$ and the $\mathbf{u} = \left\{U_0, U_1, \cdots, U_{L_g-1}\right\}$ can be retrieved by calculating the discrete Fourier transform (DFT) of $\mathbf{g}$, namely, $\mathbf{u}=\mathbf{F}\mathbf{g}$, where $\mathbf{F}$ represents the DFT matrix.

\subsubsection{Discretization of the AF}
Let $\psi_{u,v}=\psi(uT_s,v/KT_s)$ be the discretized version of the AF. We may express $\psi_{u,v}$ by replacing the integral with summation, yielding
\begin{equation}
    \psi_{u,v} = \mathbf{g}^H\mathbf{J}_{u,v}\mathbf{g},
\end{equation}
where $\mathbf{J}_{u,v}=\mathbf{J}_u\mathbf{D}_v$, 
\begin{equation}
\mathbf{J}_u = 
\begin{bmatrix}
\underbrace{0 \ldots 0}_{\text{u zeros}} & 1 & \ldots & 0 \\
0 \ldots 0 & 0 & \ddots & \vdots \\
\vdots & \vdots & \vdots & 1 \\
0 \ldots 0 & 0 & \ldots & 0
\end{bmatrix}
\end{equation}
is the time-shifting matrix, and
\begin{equation}
\mathbf{D}_v = 
\begin{bmatrix}
1 & 0 & \cdots & 0 \\
0 & e^{j2\pi v/K} & \cdots & 0 \\
\vdots & 0 & \ddots & 0 \\
0 & \cdots & 0 & e^{j2\pi v(K-1)/K} \\
\end{bmatrix}
\end{equation}
is the frequency-shifting matrix.

\subsubsection{Discretization of the ISI and Bandwidth Constraints}
Let $N_T = \left\lceil\frac{T}{T_s}\right\rceil$, $N_B = \left\lceil\frac{BT_s}{2}\right\rceil$, which denote the number of time-domain samples for each symbol duration and frequency-domain samples for the positive bandwidth of the spectrum, respectively. Then the ISI constraint may be discretized as
\begin{equation}
\chi_{nN_T,0}\leq\varepsilon_{ISI}, n=\cdots,-2,-1,1,2,\cdots,
\end{equation}
or in terms of $\mathbf{g}$ as
\begin{equation}
\mathbf{g}^H\mathbf{J}_{nN_T,0}\mathbf{g}\leq\varepsilon_{ISI}, n=\cdots,-2,-1,1,2,\cdots.
\end{equation}
Similarly, the OOBE constraint can be discretized as
\begin{equation}
\sum_{k=N_B+1}^{K-N_B-1} \left|U_k\right|^2 \leq \varepsilon_{OB},
\end{equation}
or in terms of $\mathbf{g}$ as
\begin{equation}
\mathbf{g}^{\top}\mathbf{F}^H\mathbf{E}\mathbf{F}\mathbf{g} \leq \varepsilon_{OB},
\end{equation}
where $\mathbf{E}$ is a truncation matrix that selects the high frequency components, given by
\begin{equation}
    \mathbf{E}=
    \begin{bmatrix}
        \mathbf{0}_{N_B+1,N_B+1} & \mathbf{0}_{N_B+1,K-2N_B-1} & \mathbf{0}_{N_B+1,N_B} \\
        \mathbf{0}_{K-2N_B-1,N_B+1} & \mathbf{I}_{K-2N_B-1} & \mathbf{0}_{K-2N_B-1,N_B} \\
        \mathbf{0}_{N_B,N_B+1} & \mathbf{0}_{N_B,K-2N_B-1} & \mathbf{0}_{N_B,N_B}
    \end{bmatrix}
\end{equation}

\subsubsection{Discretization of the WISL}
Define $\alpha_{n,v}=\alpha_n(v/L_gT_s)$, then the WISL can be discretized as
\begin{equation}
\begin{aligned}
\mathcal{L}_{WISL}\{\mathbf{g}\}&=\mathbb{E}\left\{\sum_{\{u,v\}\in\Theta}\sigma_{u,v}^c|\chi_{u,v}|^2\right\} \\
&= \sum_{\{u,v\}\in\Theta}\sum_{|n|<L} \alpha_{n,v}\sigma_{u,v}^c|\psi_{u+nN_T,v}|^2,
\end{aligned}
\end{equation}

\subsubsection{General ISL/WISL Minimization Problem}

With the above discretized variables, objectives and constraints at hand, we may now formulate the general WISL minimization problem as
\begin{equation}\label{general_problem}
\begin{aligned}
&\underset{\mathbf{g}}{\min} &&\mathcal{L}_{WISL}\{\mathbf{g}\} \\
&s.t. &&\mathbf{g}^H\mathbf{J}_{nN_T,0}\mathbf{g}\leq\varepsilon_{ISI}, n=\cdots,-2,-1,1,2,\cdots, \\
& &&\mathbf{g}^{\top}\mathbf{F}^H\mathbf{E}\mathbf{F}\mathbf{g} \leq \varepsilon_{OB}, \; \|\mathbf{g}\|^2 = E_g.
\end{aligned}
\end{equation}
By assigning \( \sigma^c_{u,v} = 1 \) within the region of interest in \eqref{general_problem}, the general ISL minimization problem is derived.
The general ISL/WISL minimization problem is non-convex in general. In the sequel, we will show that for the special case of minimizing the ISL/WISL of the ACF, one may formulate \eqref{general_problem} into an elegant convex QP. Moreover, by solving this problem, one may adjust $\beta$ and $\varepsilon_{ISI}$ to control the tradeoff between S\&C.

\subsection{The ACF Case Under Nyquist Constraints}
Before solving the general problems, we first investigate the ISL/WISL minimization problem for the ACF case.
\subsubsection{Discretization of the ESD}
Let us denote the non-zero part (band-limited part) of $|U(f)|^2$ by a vector $\boldsymbol{\omega}\in\mathbb{R}^{N_B+1}$. The whole sampled ESD can be constructed by
\begin{equation}
    \tilde{\boldsymbol{\omega}}=
    \begin{bmatrix}
        \boldsymbol{\omega} \\
        \mathbf{0}_{L_g-2N_B-1,1} \\
        \text{flip}(\boldsymbol{\omega}[2:N_B+1])
    \end{bmatrix}=
    \mathbf{B}\boldsymbol{\omega},
\end{equation}
where $\mathbf{B}$ is the matrix representing the linear transform $\boldsymbol{\omega}\rightarrow\tilde{\boldsymbol{\omega}}$. 
Therefore the sampled ACF sequence of $g(t)$ can be represented by the IDFT of the sampled ESD, which gives us
\begin{equation}
\boldsymbol{\psi}=\mathbf{F}^H\tilde{\boldsymbol{\omega}}=\mathbf{F}^H\mathbf{B}\boldsymbol{\omega},
\end{equation}
where $\boldsymbol{\psi}$ is the discretized ACF, which is a circularly symmetric vector. The first half of $\boldsymbol{\psi}$ is utilized for computing WISL in subsequent sections, and the $i$-th element of $\boldsymbol{\psi}$ within the first half part is defined as $\psi_{i-1}$.

\subsubsection{Discretization of the Nyquist Constraint}

Recall that $\beta$ is the roll-off factor of the pulse, such that $BT=(1+\beta)$. It holds that $N_B=(1+\beta)L_g/(2N_T)$.
Then the Nyquist condition given in (\ref{eq:freq_no_isi_constr}) can be recast as
\begin{equation}\label{linear_constraints}
\begin{aligned}
&\omega_n = N_T, 0 \leq n \leq \left[\frac{1-\beta}{1+\beta}N_B\right], \\
&\omega_n + \omega_{\left[\frac{2N_B}{1+\beta}\right]-n} = N_T, \left[\frac{1-\beta}{1+\beta}N_B\right]+1 \leq n \leq N_B.
\end{aligned}
\end{equation}
Note that \eqref{linear_constraints} already includes the energy constraint in an implicit manner, i.e., the summation of all $\omega_n$ is now a fixed constant. Moreover, by noting the fact that these constraints are linear, they can be written in the form of
\begin{equation}
    \mathbf{A}\boldsymbol{\omega}=N_T\mathbf{1}.
\end{equation}

\subsubsection{Discretization of the WISL}

For the zero-Doppler slice, $\mathcal{L}_{WISL}\{\mathbf{g}\}$ can be simplified as $\mathcal{L}_{WISL}\{\boldsymbol{\omega}\}$, which is a convex quadratic form of $\boldsymbol{\omega}$.
\begin{equation}
\begin{aligned}
\mathcal{L}_{WISL}\{\boldsymbol{\omega}\}&\approx\mathbb{E}\left\{\sum_{u\in\Theta}\sigma_u^c|\chi_{u,0}|^2\right\} \\
&= \sum_{u\in\Theta}\sum_{|n|<L} \alpha_{n,0}\sigma_{u}^c|\psi_{u+nN_T,0}|^2 \\
&= \sum_{u\in\Theta}\sum_{|n|<L} \alpha_{n,0}\sigma_{u}^c\boldsymbol{\omega}^{\top}\mathbf{T}_{u+nN_T}\boldsymbol{\omega} = \boldsymbol{\omega}^{\top}\mathbf{Q}\boldsymbol{\omega}
\end{aligned}
\end{equation}
where
\begin{equation}
\mathbf{T}_u = 
\left\{
\begin{aligned}
    &\mathbf{B}^H\mathbf{F}\mathbf{e}_u\mathbf{e}_u^H\mathbf{F}^H\mathbf{B}, \; u \leq \left\lceil \frac{L_g}{2}\right\rceil, \\
    &\mathbf{0}, \; u > \left\lceil \frac{L_g}{2}\right\rceil,
\end{aligned}
\right.
\end{equation}
and $\mathbf{Q}$ can be expressed as
\begin{equation}
    \mathbf{Q} = \sum_{u\in\Theta}\sum_{|n|<L} \alpha_{n,0}\sigma_{u}^c\mathbf{T}_{u+nN_T},
\end{equation}
which is a semidefinite matrix, and $\mathbf{e}_u$ is a selection vector with its $u$-th element being set to one while all other elements are zero.

\subsubsection{The Nyquist-ACF-QP Problem}

Based on the above framework, we may represent the WISL optimization problem for the ACF as
\begin{equation}
\label{opt:qp}
\underset{\boldsymbol{\omega}}{\min} \; \boldsymbol{\omega}^{\top}\mathbf{Q}\boldsymbol{\omega} \;\; s.t. \mathbf{A}\boldsymbol{\omega}=N_T\mathbf{1}, \; \boldsymbol{\omega}\geq \mathbf{0},
\end{equation}
which is a linearly constrained convex QP problem and may be very efficiently solved. Within this formulation, we can adjust $\beta$ to control the tradeoff between S\&C.

\subsection{Optimization Algorithms}
In this subsection, we develop efficient algorithms to solve the non-convex generic problem \eqref{general_problem} and its ACF case \eqref{opt:qp}.
\subsubsection{SCA for General WISL Minimization}
Since the general WISL objective function cannot be reformulated as a convex function of neither $\mathbf{g}$ or $\boldsymbol{\omega}$, we will have to cope with the non-convexity directly. The first challenge of the general WISL minimization is that the time-domain ISI constraint is non-convex. To deal with that, we eliminate the ISI constraint by introduce a penalty term of ISI in the objective function, with a penalty factor $\rho$, yielding
\begin{equation}\label{penalty_opt}
\begin{aligned}
&\underset{\mathbf{g}}{\min} && f(\mathbf{g}) = \mathcal{L}_{WISL}\{\mathbf{g}\} + \rho\sum_{\Abs{n}<L,n\neq 0}\mathbf{g}^H\mathbf{J}_{nN_T,0}\mathbf{g} \\
&s.t. &&\mathbf{g}^{\top}\mathbf{F}^H\mathbf{E}\mathbf{F}\mathbf{g} \leq \varepsilon_{OB}, \; \|\mathbf{g}\|^2 = E_g.
\end{aligned}
\end{equation}

The objective function $f(\mathbf{g})$ is a non-convex quadratic function of $\mathbf{g}$. To proceed, we develop a Sequential Convex Approximation (SCA) algorithm to solve \eqref{penalty_opt}, where we approximate the objective function by its first-order Taylor expansion near a given point $\mathbf{g}_k$ as
\begin{equation}
    f(\mathbf{g}) \approx f_l(\mathbf{g};\mathbf{g}_k) = f(\mathbf{g}_k) + \nabla f(\mathbf{g}_k)^\top\left(\mathbf{g}-\mathbf{g}_k\right),
\end{equation}
where 
\begin{equation}
\label{eq:pfpg}
\begin{aligned}
    \nabla f(\mathbf{g}_k) &= \sum_{\{u,v\}\in\Theta}\sum_{|n|<L} \alpha_n(v)\sigma_{u,v}^c\nabla\bar{\psi}_{u+nN_T,v}({\mathbf{g}_k}) \\
    &\quad + \rho\sum_{\Abs{n}<L,n\neq 0} \nabla\bar{\psi}_{nN_T,v}({\mathbf{g}_k}),
\end{aligned}
\end{equation}
Here $\nabla\bar{\psi}_{u,v}(\mathbf{g}_k)$, which represents the gradient of $|\psi_{u,v}|^2$ with respect to $\mathbf{g}_k$, is calculated by
\begin{equation}
\label{eq:ppsipg}
\begin{aligned}
    &\nabla\bar{\psi}_{u,v}(\mathbf{g}_k)= \\
    &= 2\left\{\Brkts{\mathbf{g}_k^{\top}\mathbf{J}_{u,v}^{\Re}\mathbf{g}_k}\mathbf{J}_{u,v}^{\Re}\mathbf{g}_k + \Brkts{\mathbf{g}_k^{\top}\Brkts{\mathbf{J}_{u,v}^{\Re}}^{\top}\mathbf{g}_k}\Brkts{\mathbf{J}_{u,v}^{\Re}}^{\top}\mathbf{g}_k\right. \\
    &\quad + \left.\Brkts{\mathbf{g}_k^{\top}\mathbf{J}_{u,v}^{\Im}\mathbf{g}_k}\mathbf{J}_{u,v}^{\Im}\mathbf{g}_k + \Brkts{\mathbf{g}_k^{\top}\Brkts{\mathbf{J}_{u,v}^{\Im}}^{\top}\mathbf{g}_k}\Brkts{\mathbf{J}_{u,v}^{\Im}}^{\top}\mathbf{g}_k\right\},
\end{aligned}
\end{equation}
where $\mathbf{J}_{u,v}^{\Re}=\Real{\mathbf{J}_{u,v}}$ and $\mathbf{J}_{u,v}^{\Im}=\Imag{\mathbf{J}_{u,v}}$.

At the $k+1$-th iteration, the SCA algorithm seeks to solve the following convex sub-problem:
\begin{equation}
\label{opt:sca}
\begin{aligned}
&\underset{\mathbf{g}}{\min} && \nabla f(\mathbf{g}_k)^{\top}\mathbf{g} \\
&s.t. &&\mathbf{g}^{\top}\mathbf{F}^H\mathbf{E}\mathbf{F}\mathbf{g} \leq \varepsilon_{OB}, \; \|\mathbf{g}\|^2 = E_g.
\end{aligned}
\end{equation}
Suppose that a solution \(\mathbf{g}_k\) has been acquired at the \(k\)-th iteration. By solving problem \eqref{opt:sca} in the \((k+1)\)-th iteration, an optimal solution \(\mathbf{g}^{\star}\) that minimizes \(f_l(\mathbf{g};\mathbf{g}_k)\) is obtained. When \(\mathbf{g}^{\star}\) is in close proximity to \(\mathbf{g}_k\) and the SCA holds, it follows that \(f(\mathbf{g}^{\star}) \leq f_l(\mathbf{g}^{\star};\mathbf{g}_k) \leq f_l(\mathbf{g}_k;\mathbf{g}_k) = f(\mathbf{g}_k)\). Although \(\mathbf{g}^{\star}\) may not be adjacent to \(\mathbf{g}_k\), the difference \(\mathbf{g}^{\star} - \mathbf{g}_k\) provides a descent direction for optimizing \(f(\mathbf{g})\). By iteratively taking small steps along the direction of \(\mathbf{g}^{\star} - \mathbf{g}_k\), it is possible to progressively obtain better solutions that minimize \(f(\mathbf{g})\).

With a properly chosen step size $t\in[0,1]$, one may get the $(k+1)$-th iteration point as
\begin{equation}
\label{opt:sca_update}
    \mathbf{g}_{k+1}=\mathbf{g}_k+t(\mathbf{g}^{\star}-\mathbf{g}_k)=(1-t)\mathbf{g}_k+t\mathbf{g}^{\star}.
\end{equation}
Since $\mathbf{g}_k,\mathbf{g}^{\star} \in \mathcal{Q}$ by the definition of convexity, we have $\mathbf{g}_{k+1} \in \mathcal{Q}$, which is a feasible solution to problem \eqref{general_problem}.

We are now ready to formally present the SCA approach to solve the problem \eqref{general_problem} in Algorithm \ref{alg:SCA}.

\begin{algorithm}[t]
\caption{SCA Algorithm for Solving \eqref{general_problem}}
\label{alg:SCA}
\begin{algorithmic}[1]
    \REQUIRE $L_g$, $N_T$, $\beta$, $\varepsilon_{OB}$, $E_g$, $\{\sigma_{u,v}^c\}$, the execution threshold $\epsilon$ and the maximum iteration number $i_{\max}$.
    \ENSURE $\mathbf{g}^{\star}$
    \STATE {Initialize $\mathbf{g}_0$ using the RRC pulse with roll-off factor $\beta$, $k=0$.}
    \REPEAT{
        \STATE Calculate the $\nabla f(\mathbf{g}_k)$ by equation \eqref{eq:pfpg} and \eqref{eq:ppsipg}.
        \STATE Solve problem \eqref{opt:sca} to obtain $\mathbf{g}^{\star}$.
        \STATE {
            Update the solution by \eqref{opt:sca_update}, where $t$ is determined by using the exact line search.
        }
        \STATE $k=k+1$.
    }
    \UNTIL {
        $\Vert \mathbf{g}_k-\mathbf{g}_{k-1}\Vert\leq\epsilon$ or $i=i_{\max}$.
    }
    \STATE $\mathbf{g}^{\star}=\mathbf{g}_k$
\end{algorithmic}
\end{algorithm}

\subsubsection{ADMM for ACF Case of WISL Minimization}

While problem \eqref{opt:qp} can be solved by the off-the-shelf numerical tools, e.g., CVX toolbox, we present a customized algorithm utilizing the ADMM to accelerate the problem-solving procedure. Numerical results indicate that this algorithm converges rapidly, typically achieving convergence in just a single iteration.

\begin{algorithm}[t]
\caption{ADMM Algorithm for Solving \eqref{opt:qp}}
\label{alg:ADMM}
\begin{algorithmic}[1]
    \REQUIRE $L_g$, $N_T$, $\beta$, $\{\sigma_{u}^c\}$, the execution threshold $\epsilon$ and the maximum iteration number $i_{\max}$.
    \ENSURE $\boldsymbol{\omega}^{\star}$
    \STATE {Initialize $\boldsymbol{\omega}_0$ using the ESD of the RRC pulse with roll-off factor $\beta$, $k=0$.}
    \STATE Calculate $\mathbf{Q}$ and $\mathbf{A}$.
    \REPEAT{
        \STATE Update $\boldsymbol{\omega}_k$, $\boldsymbol{\theta}_k$, and $\boldsymbol{\lambda}_k$ by \eqref{eq:update_omega}, \eqref{eq:update_theta}, and \eqref{eq:admm_lambda}.
        \STATE $k=k+1$.
    }
    \UNTIL {
        $\Vert \boldsymbol{\omega}_k-\boldsymbol{\omega}_{k-1}\Vert\leq\epsilon$ or $i=i_{\max}$.
    }
    \STATE $\boldsymbol{\omega}^{\star}=\boldsymbol{\omega}_k$
\end{algorithmic}
\end{algorithm}

To proceed with the ADMM framework, We first introduce an auxiliary variable, and devise the following augmented Lagrangian problem with a penalty term $\varrho>0$:
\begin{equation}
\begin{aligned}
&\underset{\boldsymbol{\omega}}{\min} &&\boldsymbol{\omega}^{\top}\mathbf{Q}\boldsymbol{\omega} + \frac{\varrho}{2}\Norm{\boldsymbol{\omega}-\boldsymbol{\theta}}^2 \\
&s.t. && \mathbf{A}\boldsymbol{\omega}=N_T\mathbf{1}, \; \boldsymbol{\theta}\geq \mathbf{0}, \; \boldsymbol{\omega}=\boldsymbol{\theta}.
\end{aligned}
\end{equation}
The augmented Lagrangian fuction can be written as
\begin{equation}
    \mathcal{L}(\boldsymbol{\omega},\boldsymbol{\theta},\boldsymbol{\lambda}) = \boldsymbol{\omega}^{\top}\mathbf{Q}\boldsymbol{\omega} + \frac{\varrho}{2}\Norm{\boldsymbol{\omega}-\boldsymbol{\theta}}^2 + \boldsymbol{\lambda}^{\top}(\boldsymbol{\omega}-\boldsymbol{\theta})
\end{equation}
The iteration format of the ADMM for \eqref{opt:qp} at $(k+1)$-th iteration can be written as
\begin{align}
    \boldsymbol{\omega}_{k+1} &= \underset{\boldsymbol{\omega}}{\text{argmin}} \; \mathcal{L}(\boldsymbol{\omega},\boldsymbol{\theta}_k,\boldsymbol{\lambda}_k) \; s.t. \; \mathbf{A}\boldsymbol{\omega}=N_T\mathbf{1}, \label{eq:admm_omega}\\
    \boldsymbol{\theta}_{k+1} &= \underset{\boldsymbol{\theta}}{\text{argmin}} \; \mathcal{L}(\boldsymbol{\omega}_{k+1},\boldsymbol{\theta},\boldsymbol{\lambda}_k) \; s.t. \; \boldsymbol{\theta}\geq \mathbf{0}, \label{eq:admm_theta} \\
    \boldsymbol{\lambda}_{k+1} &= \varrho(\boldsymbol{\omega}_{k+1}-\boldsymbol{\theta}_{k+1}) \label{eq:admm_lambda}
\end{align}

Problem \eqref{eq:admm_omega} is equivalent to 
\begin{equation}
\begin{aligned}
    \boldsymbol{\omega}_{k+1} = \underset{\boldsymbol{\omega}}{\text{argmin}} &\; \boldsymbol{\omega}^{\top}\Brkts{\mathbf{Q}+\frac{\varrho}{2}\mathbf{I}}\boldsymbol{\omega}+\Brkts{\varrho\boldsymbol{\theta}_k+\boldsymbol{\lambda}_k}^{\top}\boldsymbol{\omega} \\
    s.t. &\; \mathbf{A}\boldsymbol{\omega}=N_T\mathbf{1}
\end{aligned}
\end{equation}
According to the Karush-Kuhn-Tucker (KKT) conditions, the solution to this problem can be found by solving the following linear equation:
\begin{equation}
\label{eq:update_omega}
    \begin{bmatrix}
        2\mathbf{Q}+\varrho\mathbf{I} & \mathbf{A}^{\top} \\
        \mathbf{A} & \mathbf{0} \\
    \end{bmatrix}
    \begin{bmatrix}
        \boldsymbol{\omega}_{k+1} \\
        \tilde{\boldsymbol{\lambda}} \\
    \end{bmatrix}
    =
    \begin{bmatrix}
        -\varrho\boldsymbol{\theta}_k-\boldsymbol{\lambda}_k \\
        N_T\mathbf{1} \\
    \end{bmatrix}
\end{equation}

Moreover, problem \eqref{eq:admm_theta} is equivalent to 
\begin{equation}
\begin{aligned}
    \boldsymbol{\theta}_{k+1} = \underset{\boldsymbol{\theta}}{\text{argmin}} \; \Norm{\boldsymbol{\theta}-\Brkts{\boldsymbol{\omega}_{k+1}+\frac{1}{\varrho}\boldsymbol{\lambda}_k}} \; s.t. \; \boldsymbol{\theta}\geq\mathbf{0},
\end{aligned}
\end{equation}
which yields the following solution:
\begin{equation}
\label{eq:update_theta}
\boldsymbol{\theta}_{k+1}=\Brkts{\boldsymbol{\omega}_{k+1}+\frac{1}{\varrho}\boldsymbol{\lambda}_k}^+,
\end{equation} 
where $(\cdot)^+$ denotes the procedure of converting the negative components in the vector to zero. Based on the discussion above, the ADMM procedure for solving problem problem \eqref{opt:qp} is summarized in Algorithm \ref{alg:ADMM}.



\section{Numerical Results}
\label{sec:results}

\begin{figure*}[t]
    \centering
    \hfill
    \begin{subfigure}[t]{0.327\linewidth}
        \centering
        \includegraphics[width=\linewidth]{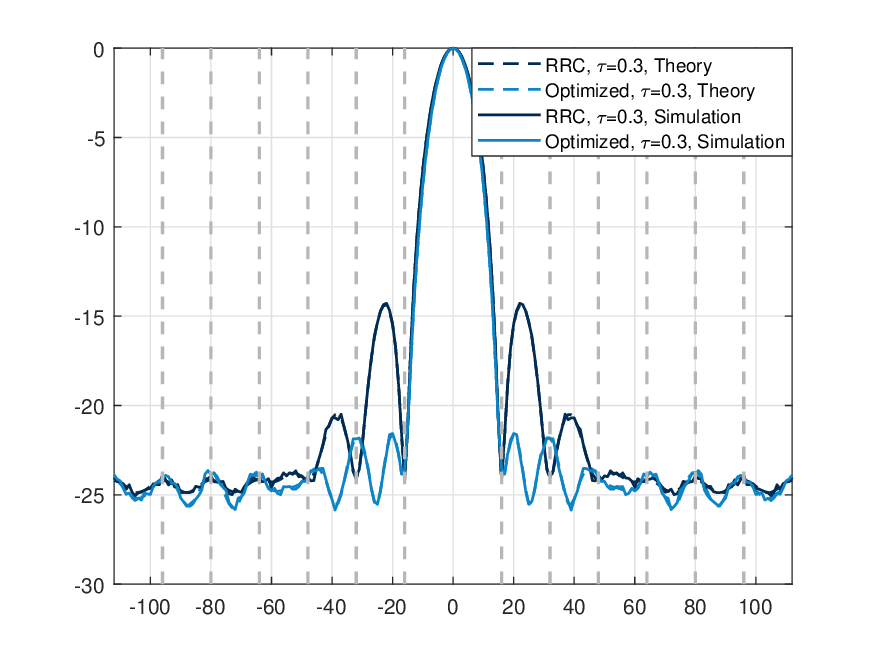}\vspace{-5pt}
        \caption{}\vspace{-5pt}
        \label{fig:islr_acf_3}
    \end{subfigure}
    \hfill
    \begin{subfigure}[t]{0.327\linewidth}
        \centering
        \includegraphics[width=\linewidth]{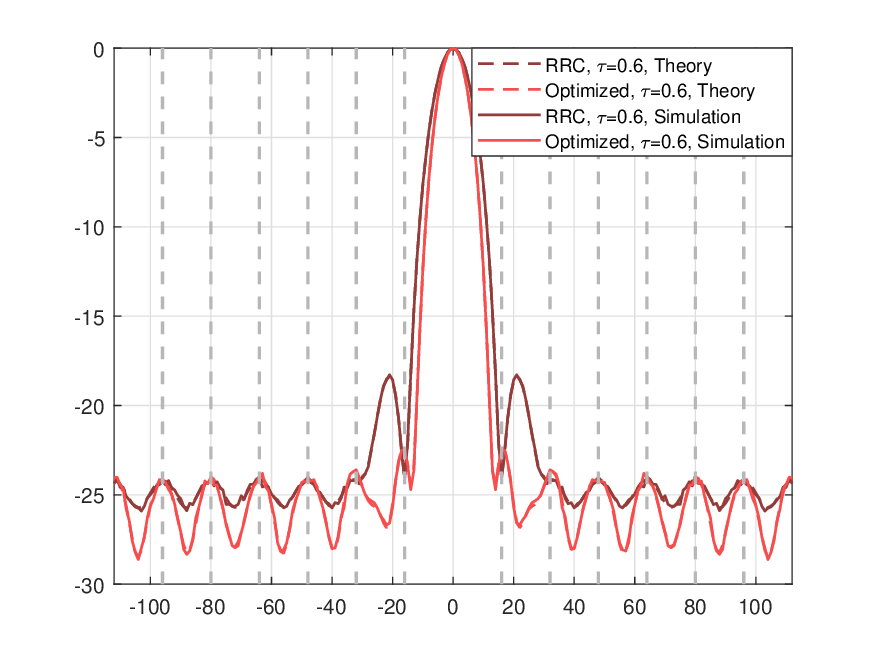}\vspace{-5pt}
        \caption{}\vspace{-5pt}
        \label{fig:islr_acf_6}
    \end{subfigure}
    \hfill
    \begin{subfigure}[t]{0.327\linewidth}
        \centering
        \includegraphics[width=\linewidth]{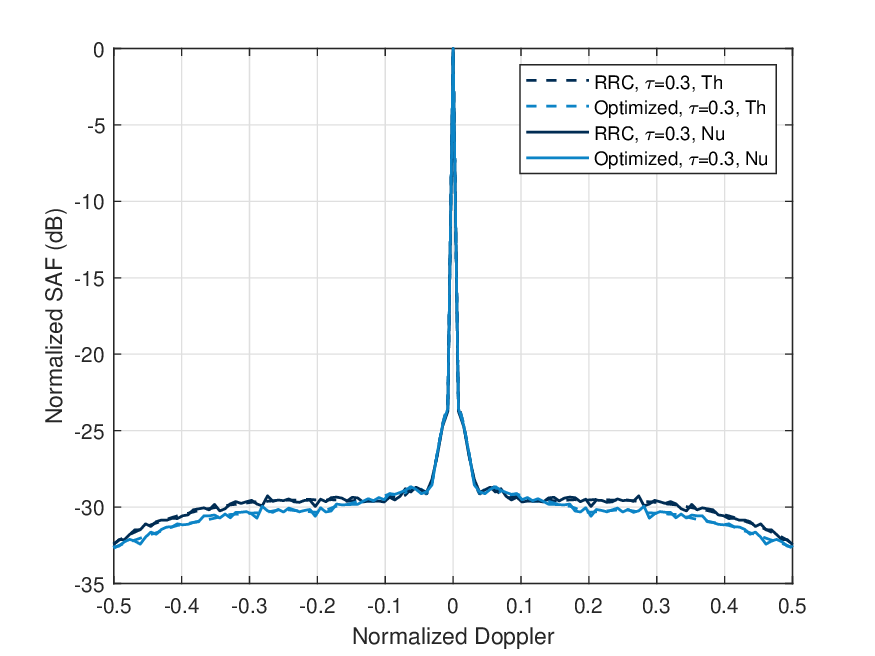}\vspace{-5pt}
        \caption{}\vspace{-5pt}
        \label{fig:islr_doppler}
    \end{subfigure}
    \caption{ {(a) Normalized SACF of the RRC and the optimized pulse by solving general ISL minimization problem. The roll-off factor $\beta=0.3$. The region of interest is set as (8m,32m) on the range domain. The grey dashed vertical lines indicate the ISI point where the SACF values are restricted to be minimal. (b) Normalized SACF of the RRC and the optimized pulse by solving general ISL minimization problem similar to the Figure \ref{fig:islr_acf_3}. The roll-off factor $\beta=0.6$. (c) Normalized SAF in the Doppler slice of the RRC and the optimized pulse by solving general ISL minimization problem. The roll-off factor $\beta=0.3$. The region of interest is set as $(0.1, 0.5)$ on the normalized Doppler domain.}}\vspace{-10pt}
    \label{fig:safs}
\end{figure*}

\begin{figure}[t]
    \centering
    \vspace{-5pt}
    \includegraphics[width=0.85\linewidth]{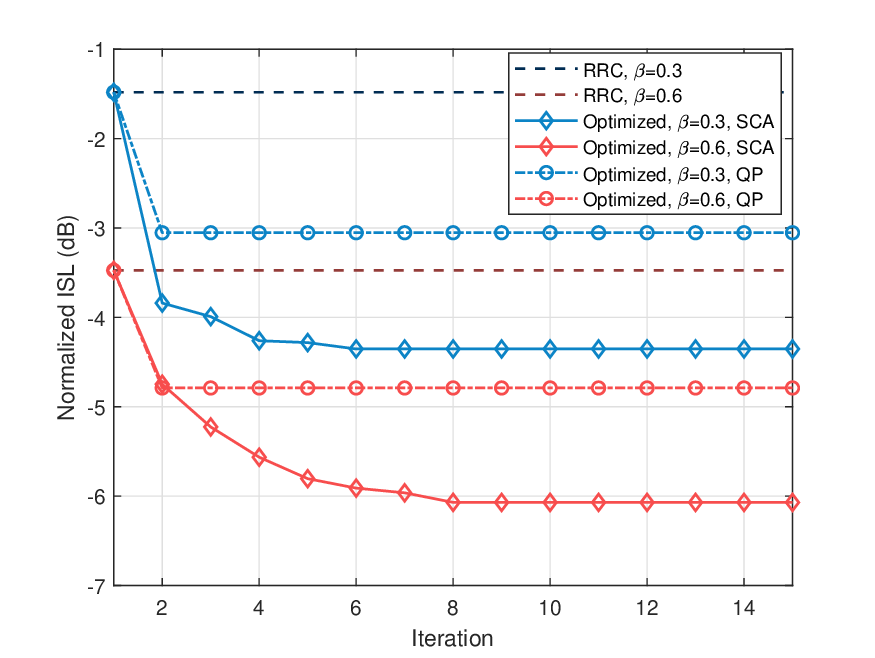}\vspace{-5pt}
    \caption{Normalized ISL versus iterations for the general ISL minimization problem solved by the SCA algorithm and the Nyquist-ACF-QP problem solved by the ADMM algorithm, with \(\beta\) set to 0.3 and 0.6.}\vspace{-5pt}
    \label{fig:islr_iter}
\end{figure}

In this section, numerical results are presented to confirm the effectiveness of the proposed pulse shaping design. The CSI configuration in the numerical simulation is considered to be range-spread and not Doppler-spread, which is consistent with the practical ISAC scenarios. The frame length $L$ is set as $256$. The sampling frequency $f_s$ (for pulse design) is established at $320$ MHz. Therefore, for $\beta=0$, the bandwidth of the signal is $20$ MHz, whereas for $\beta=1$, the bandwidth increases to $40$ MHz. The pulse length $L_g$ is set as $256$ and the number of the samples per symbol $N_T$ is set as $16$. The chosen constellation format is 16-QAM, with a $\mu_4=1.32$. In the following experiments, the value of the SAF are normalized by $\alpha_0(0)$. While other constellation types are feasible, the small value of $L(\mu_4-1)$ compared to $L^2$ in $\alpha_0(0)$ suggests that the impact of different constellations is negligible. 

\begin{figure*}[t]
    \centering
    \hfill
    \begin{subfigure}[t]{0.327\linewidth}
        \centering
        \includegraphics[width=\linewidth]{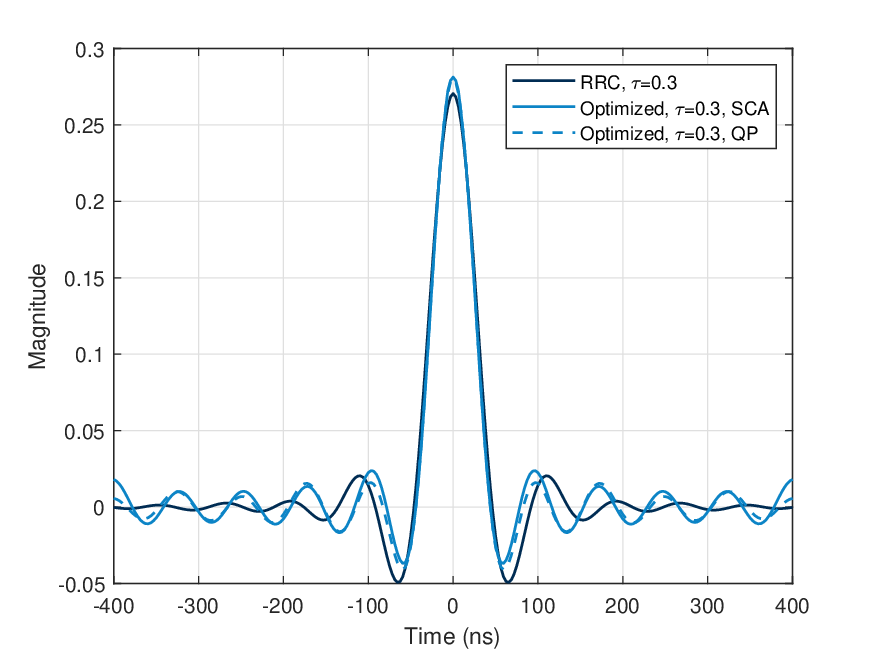}\vspace{-5pt}
        \caption{}\vspace{-5pt}
        \label{fig:islr_pulse}
    \end{subfigure}
    \hfill
    \begin{subfigure}[t]{0.327\linewidth}
        \centering
        \includegraphics[width=\linewidth]{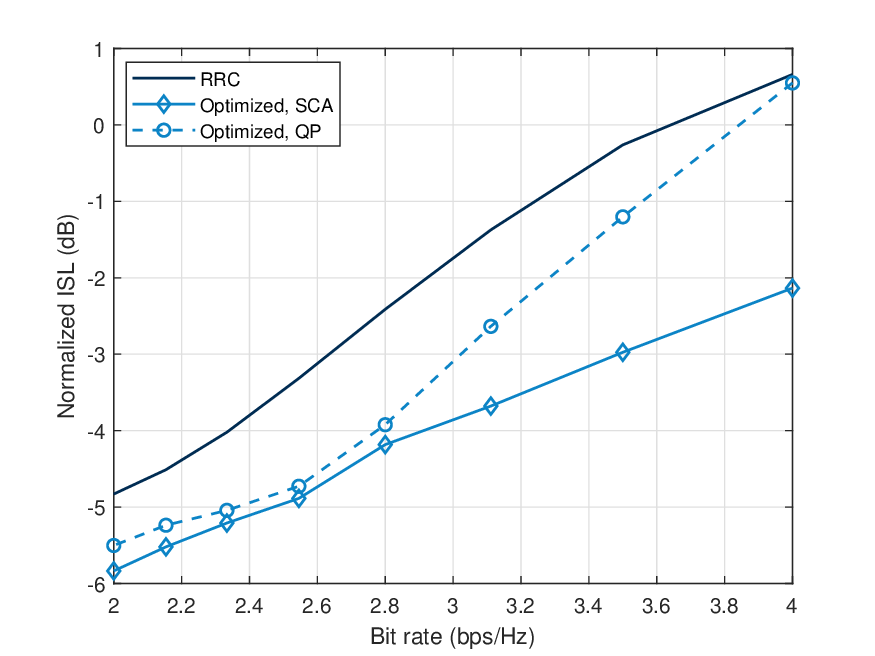}\vspace{-5pt}
        \caption{}\vspace{-5pt}
        \label{fig:islr_islrvsbr}
    \end{subfigure}
    \hfill
    \begin{subfigure}[t]{0.327\linewidth}
        \centering
        \includegraphics[width=\linewidth]{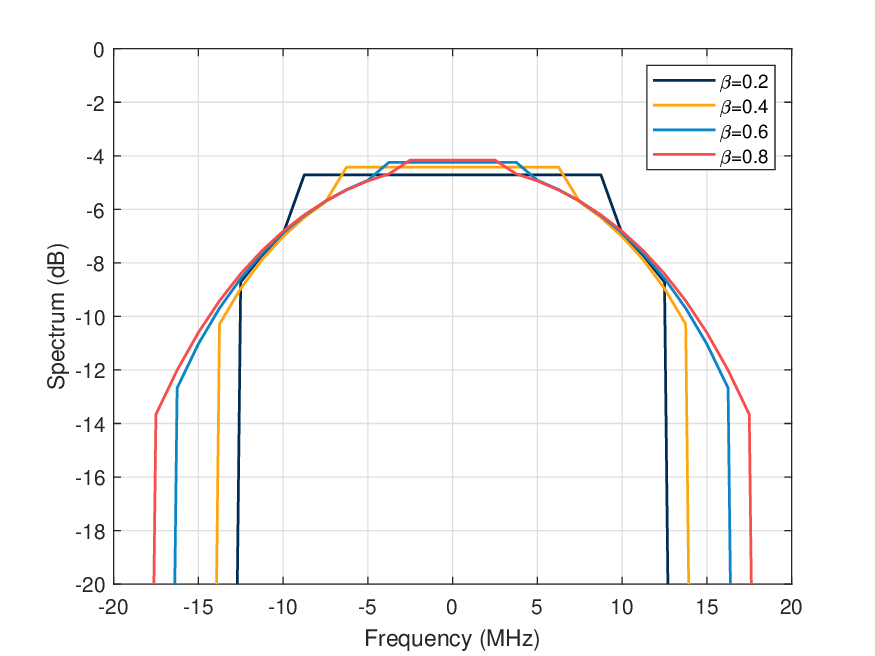}\vspace{-5pt}
        \caption{}\vspace{-5pt}
        \label{fig:islr_spectr}
    \end{subfigure}
    \caption{(a) The plot of the RRC pulse and the optimized pulse obtained by solving the general ISL minimization problem and the Nyquist-ACF-QP problem for $\beta=0.3$. (b) Normalized ISL versus bit rate for different $\beta$. The region of interest in the delay domain of the ISL is set as (8m, 32m). (c) The ESD of the RRC pulse and the optimized pulse obtained by the Nyquist-ACF-QP problem for different $\beta$.}\vspace{-10pt}
\end{figure*}

\begin{figure*}[t]
    \centering
    \hfill
    \begin{subfigure}[t]{0.327\linewidth}
        \centering
        \includegraphics[width=\linewidth]{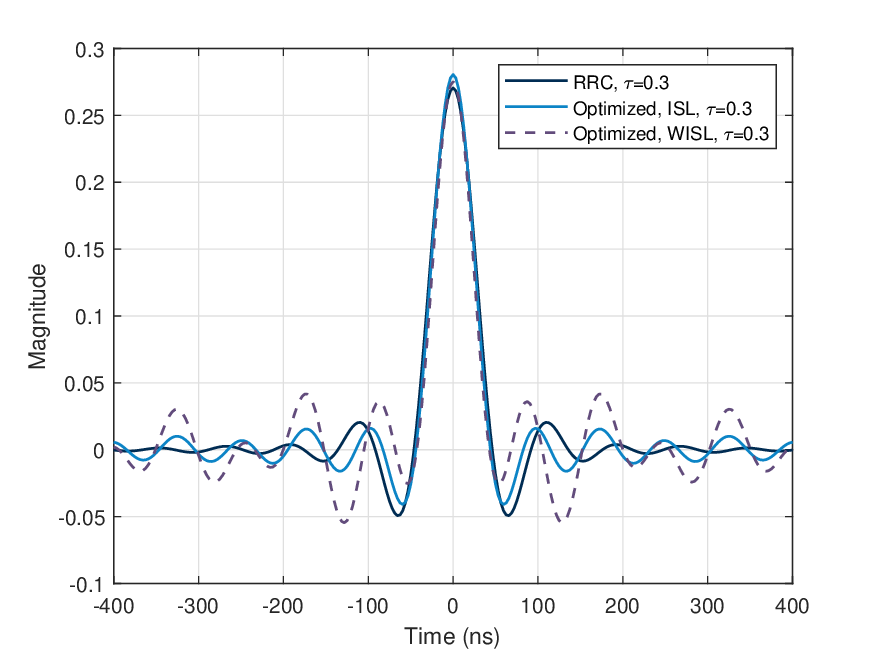}\vspace{-5pt}
        \caption{}\vspace{-5pt}
        \label{fig:wisl_pulse}
    \end{subfigure}
    \hfill
    \begin{subfigure}[t]{0.327\linewidth}
        \centering
        \includegraphics[width=\linewidth]{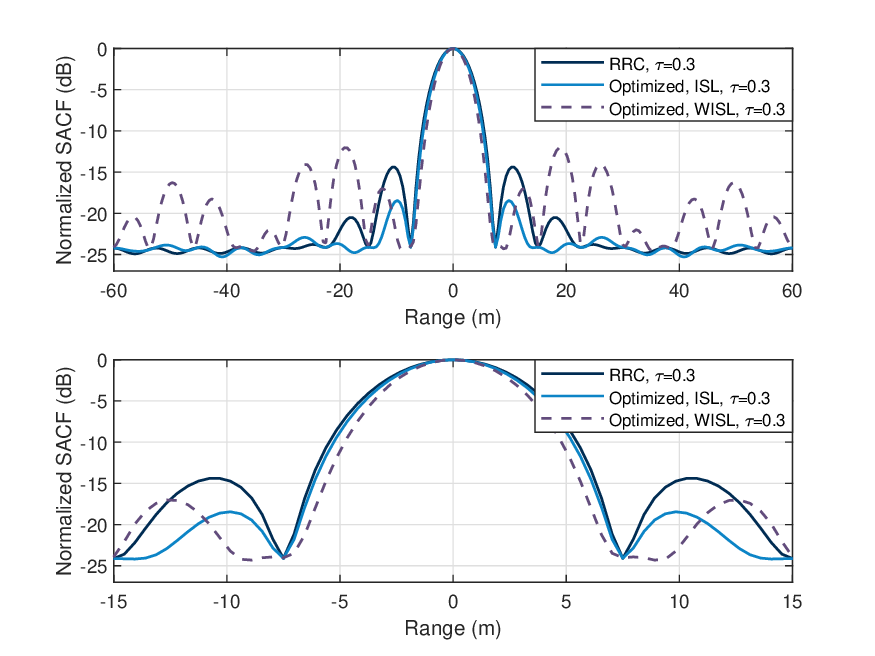}\vspace{-5pt}
        \caption{}\vspace{-5pt}
        \label{fig:wisl_acf}
    \end{subfigure}
    \hfill
    \begin{subfigure}[t]{0.327\linewidth}
        \centering
        \includegraphics[width=\linewidth]{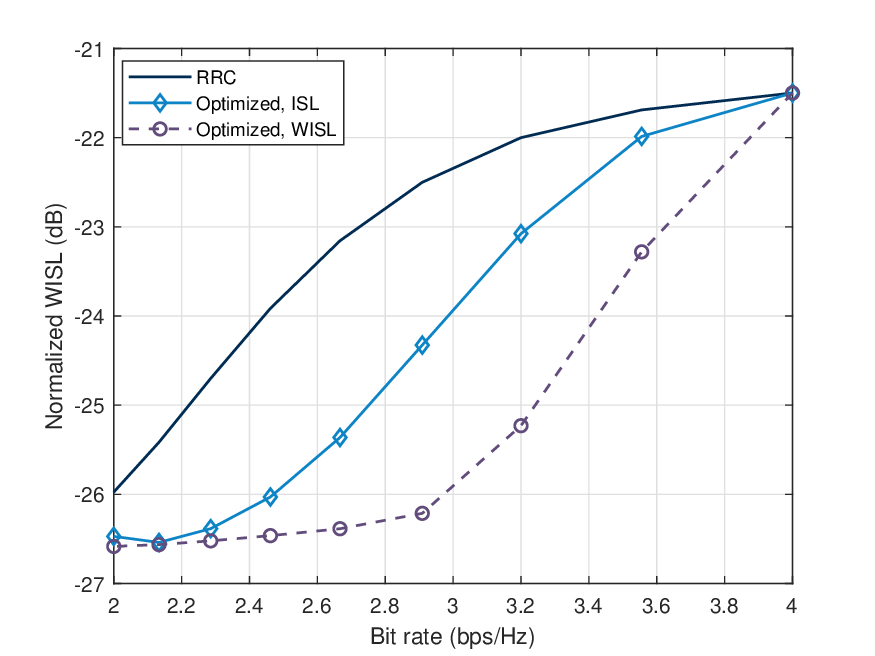}\vspace{-5pt}
        \caption{}\vspace{-5pt}
        \label{fig:wisl_wislvsbr}
    \end{subfigure}
    \caption{(a) The plot of the RRC pulse and the optimized pulse obtained by solving the Nyquist-ACF-QP problem for $\beta=0.3$. $\sigma_c(\tau)=e^{\gamma\tau}$. (b) Normalized SACF of the RRC and the optimized pulse by solving the Nyquist-ACF-QP problem for $\beta=0.3$. The region of interest is set as (8m,16m) on the range domain. (c) Normalized WISL versus the bit rate of the RRC and the optimized pulse by solving the Nyquist-ACF-QP problem by varying $\beta$. The region of interest is set as (8m,16m) on the range domain.}\vspace{-10pt}
\end{figure*}

\textit{Legends in this section:} In the legends of the Figures \ref{fig:islr_acf_3}, \ref{fig:islr_acf_6} and \ref{fig:islr_doppler}, ``Theory'' denotes that the SACF is derived from theoretical analysis, whereas ``Simulation'' indicates that the values are obtained by numerically averaging the SACF from 1000 frame signal realizations. In the legends of the Figures \ref{fig:islr_iter}, \ref{fig:islr_pulse} and \ref{fig:islr_islrvsbr}, ``SCA'' signifies that the result is obtained by solving the general ISL/WISL minimization problem, while ``QP'' denotes that the result is obtained by solving the Nyquist-ACF-QP problem. In the legends of Figures \ref{fig:wisl_pulse}, \ref{fig:wisl_acf}, and \ref{fig:wisl_wislvsbr}, ``ISL'' indicates that the results are derived under the assumption that the CSI is unavailable and $\sigma_c(\tau,\nu)$ is set to 1 within the region of interest, whereas ``WISL'' signifies that the results are obtained with the knowledge of $\sigma_c(\tau,\nu)$.

Figure \ref{fig:islr_acf_3} and \ref{fig:islr_acf_6} present the normalized squared ACF (SACF) for both the RRC pulse and the optimized pulse obtained by solving the general ISL minimization problem for \(\beta = 0.3\) and \(\beta = 0.6\) respectively. The region of interest is set as (8m, 32m) over the range domain. The SACF results are computed using theoretical values derived in \eqref{eq:theo_saf}, as well as by the numerical average of 1000 randomly generated symbol sequences. The comparison indicates that the theoretical SACF closely match their numerical counterparts. Notably, the SACF of ISAC signals with optimized pulse shaping exhibits an approximately \(6 \, \text{dB}\) reduction in the first sidelobe compared to those shaped by the RRC pulse. The SACF plot in Figure \ref{fig:islr_acf_6} exhibits similar trends to Figure \ref{fig:islr_acf_3}. 
Furthermore, Figure \ref{fig:islr_acf_6} demonstrates that the second and third sidelobe levels of the optimized SWiPS signal when \(\beta = 0.6\) are lower than those of \(\beta = 0.3\). Similarly, the second and third sidelobe levels of the RRC pulse-shaped signal when \(\beta = 0.6\) are also lower than those when \(\beta = 0.3\). This improvement in sensing performance, as will be explained later, results from a trade-off in the communication bit rate.

The grey dashed vertical lines in Figures \ref{fig:islr_acf_3} and \ref{fig:islr_acf_6} indicate the ISI points where the SACF values are constrained to be minimal. It is evident that the SACF values of the optimized pulse at the first ISI point \(G(T)\) and the second point \(G(2T)\) are slightly higher than those of the RRC pulse. This is because the ISI was not strictly constrained to be lower than that of the discrete RRC pulse. This slight compromise in ISI results in a gain in sidelobe levels compared to the optimized pulse obtained by solving the Nyquist-ACF-QP problem where ISI is strictly limited to meet the Nyquist criterion. \\
\noindent\underline{\textit{Remark:}} It should be noted that the continuous RRC pulse of infinite length theoretically has zero ISI. However, the ISI values in Figures \ref{fig:islr_acf_3} and \ref{fig:islr_acf_6} are not completely zero because we are considering a discrete RRC pulse of finite length.

Figure \ref{fig:islr_doppler} illustrates the squared zero-delay slice of the AF for the ISAC signal shaped by both the RRC pulse and the optimized pulse, which are attained by solving the general ISL minimization problem for \(\beta = 0.3\) when the region of interest is set as \((0.1, 0.5)\) on the normalized Doppler domain. While Figures \ref{fig:islr_acf_3} and \ref{fig:islr_acf_6} demonstrate the feasibility of optimizing the ranging sidelobe levels, Figure \ref{fig:islr_doppler} reveals that the performance gain on the Doppler slice is marginal.
We conclude that the limited improvement in Doppler ISL performance is attributable to the inherent characteristics of single-carrier SWiPS. Significant enhancement in joint range-Doppler ISL may be possible in other signaling formats, e.g., OFDM, which could be addressed in future research.

Figure \ref{fig:islr_iter} demonstrates the normalized ISL values as a function of iterations for both the general ISL minimization problem, solved by the SCA algorithm, and the Nyquist-ACF-QP problem, solved by the ADMM algorithm, with \(\beta\) values of 0.3 and 0.6. It is evident that the final normalized ISL achieved by solving the general ISL minimization problem is lower than that of the Nyquist-ACF-QP problem. As discussed earlier, this improvement is due to a slight trade-off in terms of the ISI. Additionally, it is noticeable that the ADMM algorithm converges in just a single iteration, whereas the SCA algorithm requires approximately 6 to 8 iterations to converge.

Figure \ref{fig:islr_pulse} displays the RRC pulse and the optimized pulse obtained by solving the general ISL minimization problem and the Nyquist-ACF-QP problem for \(\beta = 0.3\). It is evident that the pulse derived from the general ISL minimization problem closely aligns with that obtained from the Nyquist-ACF-QP problem. Additionally, the decay rate of the optimized pulse is slower than that of the RRC pulse.

Figure \ref{fig:islr_islrvsbr} illustrates the tradeoff between ISL and bit rate by adjusting \(\beta\). As \(\beta\) increases, the ISL is also on the rise, as depicted in the figure. At \(\beta = 0\), the maximum bit rate is achieved with the sinc pulse, which is the only pulse meeting the Nyquist criterion, resulting in no improvement in the S\&C tradeoff even when shaped by the optimized pulse obtained from solving the Nyquist-ACF-QP problem. It is observed that the smaller the \(\beta\), the greater the ISL gain between solving the general ISL minimization problem and the Nyquist-ACF-QP problem. This is because with smaller \(\beta\), the bandwidth constraint becomes tighter, leading to fewer solutions that satisfy the Nyquist criterion. However, the general ISL relaxes the Nyquist criterion to an $\varepsilon$-ISI constraint, creating more feasible solutions when \(\beta\) is smaller.\\
\noindent\underline{\textit{Remark:}} The bit rate is calculated as \( 4/(TB) = 4/(\beta + 1) \) bps/Hz, where the factor of 4 arises from the 16-QAM modulation, with each symbol carrying 4 bits. Please note that the bit rate does not equal to the capacity. For pulses obtained by solving the general WISL minimization problem, where the ISI is small but not necessarily zero, the capacity may be slightly lower than that achieved by the Nyquist-ACF-QP problem. This exemplifies the tradeoff between S\&C performance.

Figure \ref{fig:islr_spectr} displays the ESD of the RRC pulse and the optimized pulse obtained by solving the Nyquist-ACF-QP problem for various values of \(\beta\). As \(\beta\) increases, the bandwidth restriction for the problem is relaxed. It is evident that the optimal pulse for larger \(\beta\) does not concentrate most of its energy in the low-frequency range. This observation is consistent with the findings in \cite{xiong2024snr}, which demonstrated that when the noise is negligible (i.e., in the high signal-to-noise ratio (SNR) regime), the optimal sensing pulse tends to allocate more of its energy to the high-frequency components.

In Figures \ref{fig:wisl_pulse}, \ref{fig:wisl_acf} and \ref{fig:wisl_wislvsbr}, the function \(\sigma_c(\tau)\) for WISL minimization where the CSI is available is set to decay exponentially with respect to \(\tau\), specifically, \(\sigma_c(\tau) = e^{\gamma\tau}\), where \(\gamma\) is the decay factor. Figure \ref{fig:wisl_pulse} displays the RRC pulse and the optimized pulse obtained by solving the Nyquist-ACF-QP problem for \(\beta = 0.3\) with and without CSI. It can be observed that the pulse optimized for WISL differs significantly from the pulse optimized for ISL.

Figure \ref{fig:wisl_acf} demonstrates the normalized SACF for the RRC pulse and the optimized pulse obtained by solving the Nyquist-ACF-QP problem for \(\beta = 0.3\) with and without CSI. Compared to the pulse obtained without CSI, the sidelobe levels of the pulse obtained with the statistical information of \(\sigma_c(\tau)\) are significantly higher. However, upon closer inspection around the mainlobe, it is evident that the SACF value near the mainlobe is minimized. This is because the weight \(\sigma_c(\tau)\) near the mainlobe is much higher than in regions far from the mainlobe. Consequently, the optimal solution to the WISL minimization problem trades sidelobe levels for a lower SACF value around the mainlobe to maintain a low WISL. This suggests that the optimal pulse patterns with and without CSI considerably differ from each other. Figures \ref{fig:wisl_wislvsbr} displays the normalized WISL versus the bit rate for the RRC pulse and the optimized pulse obtained by solving the Nyquist-ACF-QP problem for different $\beta$ with and without CSI. Even without CSI, the Nyquist-ACF-QP problem effectively reduces the WISL. However, a performance gap remains between the cases with and without CSI, underscoring the importance of considering CSI.

\begin{figure*}[t]
    \centering
    \hfill
    \begin{subfigure}[b]{0.327\linewidth}
        \centering
        \includegraphics[width=\linewidth]{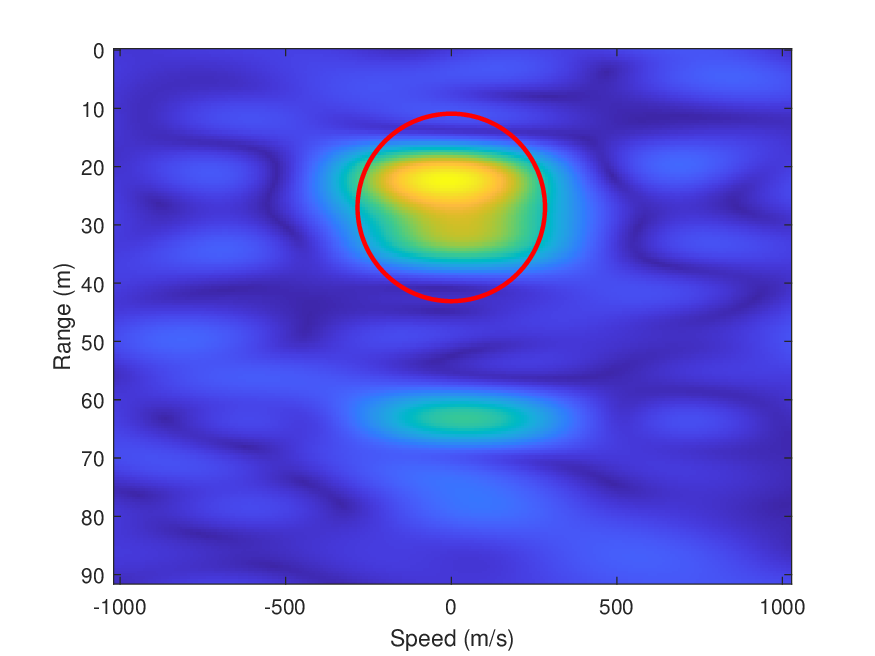}\vspace{-5pt}
        \caption{}\vspace{-5pt}
        \label{fig:rdmap_rrc}
    \end{subfigure}
    \hfill
    \begin{subfigure}[b]{0.327\linewidth}
        \centering
        \includegraphics[width=\linewidth]{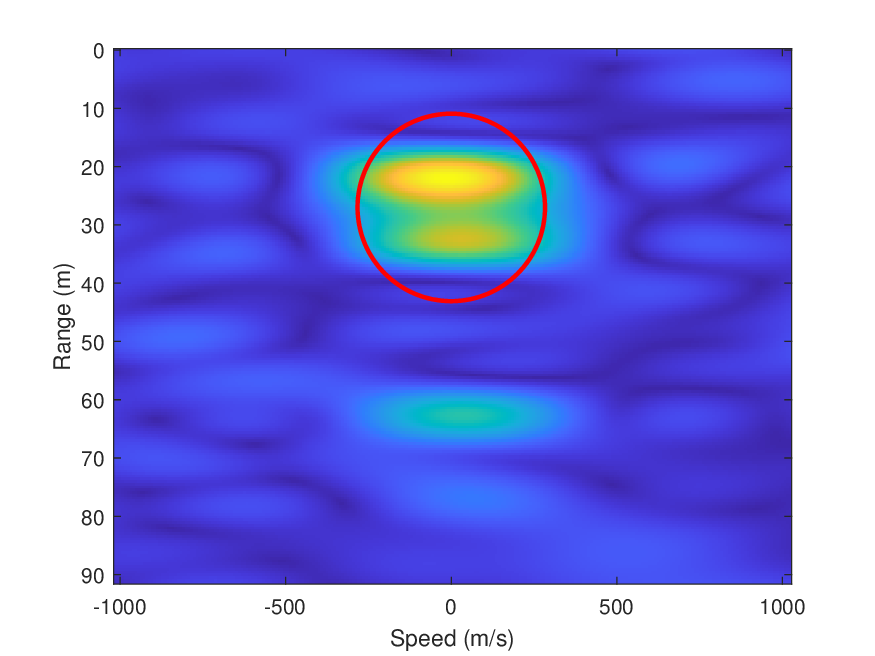}\vspace{-5pt}
        \caption{}\vspace{-5pt}
        \label{fig:rdmap_opt}
    \end{subfigure}
    \hfill
    \begin{subfigure}[b]{0.327\linewidth}
        \centering
        \includegraphics[width=\linewidth]{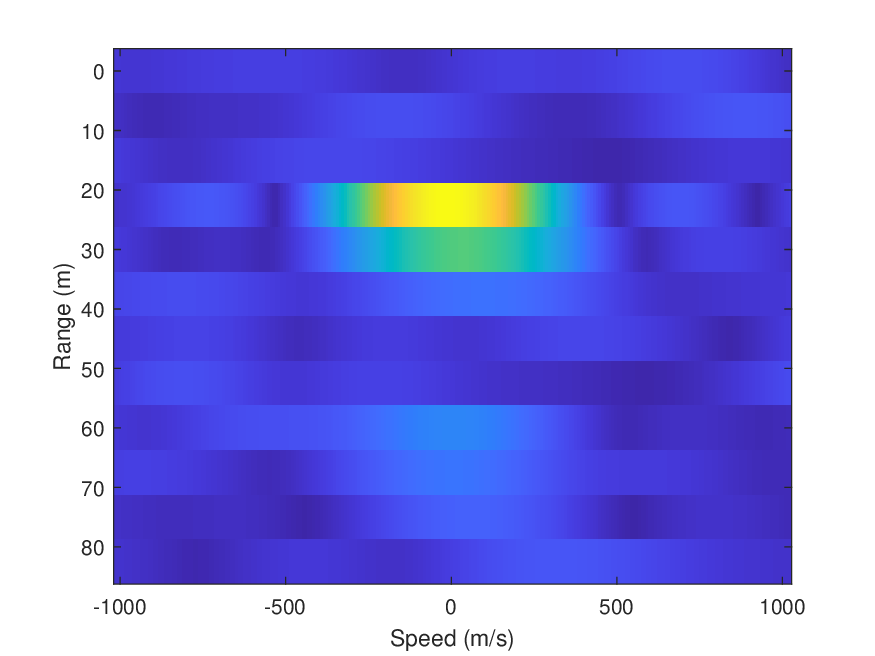}\vspace{-5pt}
        \caption{}\vspace{-5pt}
        \label{fig:rdmap_sym}
    \end{subfigure}
    \caption{Range-Doppler map of (a) the RRC pulse, (b) the optimized pulse, and (c) the symbol sequence. The red circle in (a) and (b) highlights the area where improvements have been made. In (c), the computation of the Range-Doppler map utilizes only the symbol sampling points, with the sampling duration equal to the symbol duration.}\vspace{-10pt}
    \label{fig:rdmap}
\end{figure*}

\begin{figure}[t]
    \centering
    \vspace{-5pt}
    \includegraphics[width=0.85\linewidth]{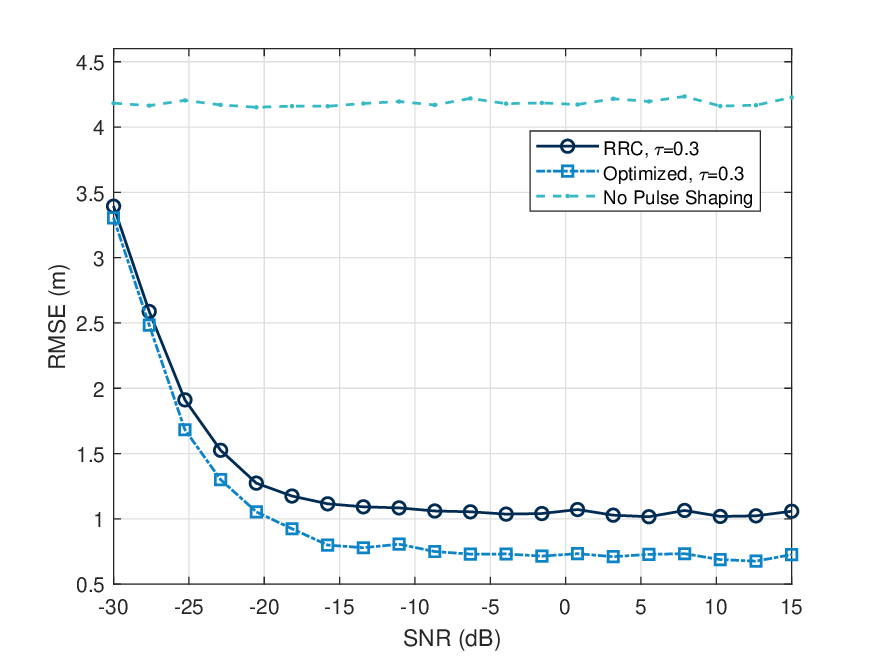}\vspace{-5pt}
    \caption{Range RMSE versus SNR for the RRC and the optimized pulse obtained by solving the Nyquist-ACF-QP problem without CSI and with $\beta=0.3$, compared to using only the symbol sequence without considering pulse shaping.}\vspace{-10pt}
    \label{fig:islr_mse}
\end{figure}

Figure \ref{fig:rdmap} presents the Range-Doppler maps measured by ISAC signals with RRC pulse, the optimized pulse, and the symbol sequence without pulse shaping, as depicted in Figures \ref{fig:rdmap_rrc}, \ref{fig:rdmap_opt}, and \ref{fig:rdmap_sym}, respectively. The experimental setup includes three targets positioned at 22m, 32m, and 63m, with reflection coefficients of 1, 0.8, and 0.5, respectively. As illustrated in Figure \ref{fig:rdmap_rrc}, the peak corresponding to the second target is obscured by the peak of the first target when shaped by the RRC pulse. In contrast, this occlusion is absent in Figure \ref{fig:rdmap_opt}, demonstrating the superior resolution of the optimized pulse. Furthermore, Figure \ref{fig:rdmap_sym} underscores the critical importance of pulse shaping, as the resolution of the Range-Doppler map is significantly reduced when only the symbol sequence is considered, resulting in the failure of the target detection.

Figure \ref{fig:islr_mse} illustrates the range Root Mean Square Error (RMSE) as a function of Signal-to-Noise Ratio (SNR) for the ISAC signal shaped by the RRC pulse and the optimized pulse, where the latter is attained by solving the Nyquist-ACF-QP problem without CSI for $\beta = 0.3$. The ranging results are compared to a scenario utilizing solely the symbol sequence without considering pulse shaping. The experiment considers two targets located within the range intervals [18m, 27m] and [32m, 41m], with reflection coefficients set at 1 and 0.8, respectively. The results demonstrate that relying exclusively on the sampled symbol sequence for range localization results in imprecise range estimation due to insufficient resolution and the off-grid nature of the target ranges. Additionally, Figure \ref{fig:islr_mse} indicates that the optimized pulse achieves superior ranging accuracy when targets are in close proximity. This is particularly evident when the distance between the targets falls within the first sidelobe region of the pulse, where the SACF values of the optimized pulses are significantly lower than those of the RRC pulses.

\section{Conclusion}

In this paper, we introduce a novel SWiPS design for single-carrier ISAC system, addressing the challenges of integrating sensing-oriented performance metrics into the SWiPS framework. We established the relationship between the AF of the frame signal and that of the pulse, and analyzed the statistical properties of the AF. We formulated the ISL and WISL to evaluate the ambiguity characteristics and presented an advanced sensing input-output model that incorporates CSI, providing a more comprehensive performance indicator. To optimize the SWiPS design, we developed algorithms utilizing the SCA and ADMM to solve the ISL/WISL minimization problem. Our numerical results validated the theoretical analysis, and demonstrated that the proposed SWiPS design significantly reduces ranging ISL/WISL, compared to traditional RRC pulse shaping, while maintaining communication bit rate and desired signal properties. Our future work may involve extending the concept of SWiPS design to multi-carrier systems, which has the potential to improve both range and Doppler ambiguity characteristics, as opposed to the marginal improvements observed in the Doppler domain in this paper.

\newcounter{appendixcounter}
\renewcommand\appendix{\par
    \setcounter{appendixcounter}{1} 
    \renewcommand{\thesection}{\Alph{appendixcounter}} 
    \titleformat{\section} 
        {\centering\sc} 
        {Appendix \Alph{appendixcounter}:} 
        {0.5em} 
        {\addtocounter{appendixcounter}{1}} 
}

\appendix

\begin{figure*}[t]
\centering
\begin{equation}
\begin{aligned}
\Expct{\chi_s(\tau,\nu)\chi_c^*(\tau,\nu)} &= \Expct{\left(\sum_{n'=0}^{L-1}\left|s_{n'}\right|^2 \psi_{n',n'}(\tau,\nu)\right) \left(\sum_{n=0}^{L-1} \sum_{m=0, m \neq n}^{L-1} s_n s_m^* \psi_{n,m}(\tau,\nu)\right)^*} \\
&= \sum_{n'=0}^{L-1}\sum_{n=0}^{L-1}\sum_{m=0, m \neq n}^{L-1} \Expct{\Abs{s_{n'}}^2 s_n^* s_m} \psi_{n',n'}(\tau,\nu) \psi_{n,m}^*(\tau,\nu) = 0. \\
\end{aligned}
\label{eq:corr_sc_expct}
\end{equation}
\begin{equation}
\begin{aligned}
    \Var{\chi_s(\tau, \nu)} &= \Expct{\Abs{\chi_s(\tau, \nu)}^2} - \Abs{\Expct{\chi_s(\tau, \nu)}}^2 = \Expct{\Abs{\sum_{n=0}^{L-1}\left|s_n\right|^2 \psi_{n,n}(\tau,\nu)}} - \Abs{\psi(\tau, \nu)\sum_{n=0}^{L-1} e^{j 2 \pi n \nu T}}^2 \\
    &= \Abs{\psi(\tau,\nu)}^2\sum_{n=0}^{L-1}\sum_{m=0}^{L-1} \Expct{\Abs{s_n}^2\Abs{s_m}^2} e^{-j2\pi (n-m)\nu T} - \Abs{\psi(\tau,\nu)}^2\sum_{n=0}^{L-1}\sum_{m=0}^{L-1} e^{-j2\pi (n-m)\nu T} \\
    &= \Abs{\psi(\tau,\nu)}^2\sum_{n=0}^{L-1}\sum_{m=0}^{L-1} \left(\Expct{\Abs{s_n}^2\Abs{s_m}^2}-1\right) e^{-j2\pi (n-m)\nu T} = \Abs{\psi(\tau,\nu)}^2\sum_{n=0}^{L-1} \left(\Expct{\Abs{s_n}^4}-1\right)
\end{aligned}
\label{eq:corr_chis_app}
\end{equation}
\begin{equation}
\begin{aligned}
    \Var{\chi_c(\tau, \nu)} &= \Expct{\Abs{\chi_c(\tau, \nu)}^2} = \Expct{\Abs{\sum_{n=0}^{L-1} \sum_{m=0, m \neq n}^{L-1} s_n s_m^* \psi_{n,m}(\tau,\nu)}^2} \\
    &= \sum_{n=0}^{L-1}\sum_{m=0, m \neq n}^{L-1}\sum_{n'=0}^{L-1}\sum_{m'=0, m' \neq n'}^{L-1} \Expct{s_n s_m^* s_{n'}^* s_{m'}} \psi_{n,m}(\tau,\nu)\psi_{n',m'}^*(\tau,\nu)
    = \sum_{n=0}^{L-1}\sum_{m=0, m \neq n}^{L-1} \Abs{\psi_{n,m}(\tau,\nu)}^2
\end{aligned}
\label{eq:corr_chic_app}
\end{equation}
\begin{equation}
\begin{aligned}
    \mathbb{E}\left[|X(\tau, \nu)|^2 | s_{0:L-1}\right] &= \iint\iint \Expct{\alpha(\lambda, \mu)\alpha^*(\lambda', \mu')} \chi(\tau - \lambda, \nu - \mu)\chi^*(\tau - \lambda', \nu - \mu') d\lambda d\mu d\lambda' d\mu' \\
    &= \iint\iint \sigma(\lambda, \mu)\delta(\lambda - \lambda')\delta(\mu - \mu') \chi(\tau - \lambda, \nu - \mu)\chi^*(\tau - \lambda', \nu - \mu') d\lambda d\mu d\lambda' d\mu' \\
    &= \iint \sigma(\lambda, \mu) \Abs{\chi(\tau - \lambda, \nu - \mu)}^2 d\lambda d\mu d\lambda' d\mu'
\end{aligned}
\label{eq:wssus}
\end{equation}
\hrule
\vspace{-15pt}
\end{figure*}



\section{Proof of Proposition \ref{prop:chi}}
\label{sec:app_B}

Recall that in Assumption \ref{assump:independent} we have $\Expct{s_n s_m^*}=0$ for all instances where $n\neq m$ and that in Assumption \ref{assump:identical} we have $\Expct{\Abs{s_n}^2}=1,\forall n$, thus we have
\begin{align}
&\begin{aligned}
    \mathbb{E}\{\chi_s(\tau, \nu)\} = \sum_{n=0}^{L-1} \psi_{n,m}(\tau,\nu) = \psi(\tau, \nu)\sum_{n=0}^{L-1} e^{j 2 \pi n \nu T} ,
\end{aligned} \\
&\begin{aligned}
    \Expct{\chi_c(\tau,\nu)} &= \Expct{\sum_{n=0}^{L-1} \sum_{m=0, m \neq n}^{L-1} s_n s_m^* \psi_{n,m}(\tau,\nu)}\\
    &= \sum_{n=0}^{L-1} \sum_{m=0, m \neq n}^{L-1} \Expct{s_n s_m^*} \psi_{n,m}(\tau,\nu) = 0.
\end{aligned}
\end{align}

Recall that in Assumption \ref{assump:symmetric} we have $\Expct{s_n}=0,\forall n$. Thus for any tuple $\left(n',n,m\right)$ where $n\neq m$, we have that
\begin{equation}
\begin{aligned}
&\Expct{\Abs{s_{n'}}^2 s_n^* s_m}=\\
&\left\{
\begin{aligned}
    &\Expct{\Abs{s_{n}}^2 s_n^*}\Expct{s_m}=0, &&n'=n\neq m, \\
    &\Expct{\Abs{s_{m}}^2 s_m}\Expct{s_n}^*=0, &&n'=m\neq n, \\
    &\Expct{\Abs{s_{n'}}^2}\Expct{s_n}^*\Expct{s_m}=0, &&n'\neq m, n'\neq n.
\end{aligned}
\right.
\end{aligned}
\end{equation}
Then according to \eqref{eq:corr_sc_expct} we have $\Expct{\chi_s(\tau,\nu)\chi_c^*(\tau,\nu)}=0$.

\section{Proof of Proposition \ref{prop:chisc_var}}
\label{sec:app_C}

By the independence between different symbols, we have
\begin{equation}
    \Expct{\Abs{s_n}^2\Abs{s_m}^2} = \Expct{\Abs{s_n}^2}\Expct{\Abs{s_m}^2} = 1, \forall n\neq m.
\end{equation}
Given \eqref{eq:corr_chis_app} and the condition $\Expct{\Abs{s_n}^4}=\mu_4,\forall n$, we conclude that \eqref{eq:chis_var} holds. Recall that in Assumption \ref{assump:symmetric} we have $\Expct{s_n^2}=0,\forall n$. Thus for any tuple $\left(n,m,n',m'\right)$ where $n\neq m, n'\neq m$, we have
\begin{equation}
\begin{aligned}
&\Expct{s_n^* s_m s_{n'}^* s_{m'}}=\\
&\left\{
\begin{aligned}
    &\Expct{\Abs{s_{n}}^2}\Expct{\Abs{s_{m}}^2}=1, &&n=n',m=m',n\neq m, \\
    &\Expct{s_{n}^2}\Expct{s_{m}^2}=0, &&n=m',m=n',n\neq m, \\
\end{aligned}
\right.
\end{aligned}
\end{equation}
Then according to \eqref{eq:corr_chic_app} and the fact that $\Abs{\psi_{n,m}(\tau,\nu)}^2=\Abs{\psi(\tau+(m-n)T,\nu)}^2$, we have
\begin{equation}
\begin{aligned}
    &\Var{\chi_c(\tau, \nu)} = \sum_{n=0}^{L-1}\sum_{m=0, m \neq n}^{L-1} \Abs{\psi(\tau+(m-n)T,\nu)}^2 \\
    &\; = \sum_{n=1}^{L-1}(L-n)\left(|\psi(\tau-nT, v)|^2 + |\psi(\tau+nT, v)|^2\right).
\end{aligned}
\end{equation}
Thus we conclude that \eqref{eq:chic_var} holds. The variance of the AF can be expressed as 
\begin{equation}
\begin{aligned}
    &\Var{\chi(\tau,\nu)}=\underbrace{\Expct{\Abs{\chi_s(\tau,\nu)}^2}-\Abs{\Expct{\chi_s(\tau,\nu)}}^2}_{\Var{\chi_s(\tau, \nu)}} \\
    &\quad + \underbrace{\Expct{\Abs{\chi_c(\tau,\nu)}^2}-\Abs{\Expct{\chi_c(\tau,\nu)}}^2}_{\Var{\chi_c(\tau, \nu)}} \\
    &\quad + \underbrace{2\Real{\Expct{\chi_s(\tau,\nu)\chi_c^*(\tau,\nu)}-\Expct{\chi_s(\tau,\nu)}\Expct{\chi_c^*(\tau,\nu)}}}_{\text{cross term}}.
\end{aligned}
\end{equation}
From \eqref{eq:zero_chic} and \eqref{eq:zero_corr_sc} we have that the cross term is equal to $0$, which gives us \eqref{eq:var_addition}.

\section{Proof of Proposition \ref{prop:wssus}}
\label{sec:app_E}

The conditional (given the symbols) mean square value of $X(\tau, \nu)$, under the WSSUS assumption, is writen as \eqref{eq:wssus}, whose value at $(0,0)$ is:
\begin{equation}
    \mathbb{E}\left[|X(0,0)|^2 | s_{0:L-1}\right] = \iint \sigma(\lambda, \mu)|\chi(-\lambda, -\mu)|^2 d\lambda d\mu.
\end{equation}
By using the fact that $|\chi(-\lambda, -\mu)|=|\chi(\lambda, \mu)|$ and averaging over the symbols, we thus obtain
\begin{equation}
    \mathbb{E}[|X(0,0)|^2] = \iint \sigma(\lambda, \mu)\mathbb{E}\left[|\chi(\lambda, \mu)|^2\right] d\lambda d\mu.
\end{equation}
\vspace{-20pt}

\bibliographystyle{ieeetr}
\bibliography{ref}




\end{document}